\documentclass[10pt]{article}

\usepackage[superscript,sort]{cite}

\usepackage{ifthen,ifpdf}
\ifpdf
	\usepackage[pdftex]{hyperref}	\hypersetup{colorlinks=true,linkcolor=black,citecolor=black,urlcolor=blue,backref=page,bookmarks=true,breaklinks=true,plainpages=false }
	\usepackage[pdftex]{graphicx}
	\usepackage[usenames,pdftex]{color}
\else
	\usepackage[colorlinks=true,breaklinks=true,plainpages=false]{hyperref}
	\usepackage{graphicx}
	\usepackage[usenames]{color}
\fi

\usepackage{amsthm,amscd,amsxtra,amsfonts,amsmath,amssymb,multirow}
\usepackage{wrapfig}
\usepackage[footnotesize]{caption}
\usepackage[tiny,compact]{titlesec}
\usepackage[textwidth=0.8in,textsize=footnotesize]{todonotes}
\usepackage{algorithm,algorithmic,extarrows}

\newtheorem{thm}{Theorem}
\newtheorem{defn}{Definition}
\newtheorem{lem}{Lemma}[thm]

\newtheorem{prop}{Proposition}[thm]

\begin{document}

\title{Weighted persistent homology for biomolecular data analysis}

\author{
Zhenyu Meng$^{1,2}$,
D Vijay Anand$^1$,
Yunpeng Lu$^2$, Jie Wu $^4$, and
Kelin Xia$^{1,3}$\footnote{ Address correspondences  to Kelin Xia. E-mail:xiakelin@ntu.edu.sg}
\\
$^1$Division of Mathematical Sciences, School of Physical and Mathematical Sciences, \\
Nanyang Technological University, Singapore 637371\\
$^2$Division of Chemistry and Biological Chemistry, School of Physical and Mathematical Sciences, \\
Nanyang Technological University, Singapore 637371\\
$^3$School of Biological Sciences, Nanyang Technological University, Singapore 637371\\
$^4$Department of Mathematics, National University of Singapore, Singapore
}

\date{\today}
\maketitle

\begin{abstract}
In this paper, we systematically review weighted persistent homology (WPH) models and their applications in biomolecular data analysis. Essentially, the weight value, which reflects physical, chemical and biological properties, can be assigned to vertices (atom centers), edges (bonds), or higher order simplexes (cluster of atoms), depending on the biomolecular structure, function, and dynamics properties. Further, we propose the first localized weighted persistent homology (LWPH). Inspired by the great success of element specific persistent homology (ESPH), we do not treat biomolecules as an inseparable system like all previous weighted models, instead we decompose them into a series of local domains, which may be overlapped with each other. The general persistent homology or weighted persistent homology analysis is then applied on each of these local domains. In this way, functional properties, that are embedded in local structures, can be revealed. Our model has been applied to systematically studying DNA structures. It has been found that our LWPH based features can be used to successfully discriminate the A-, B-, and Z-types of DNA. More importantly, our LWPH based PCA model can identify two configurational states of DNA structure in ion liquid environment, which can be revealed only by the complicated helical coordinate system. The great consistence with the helical-coordinate model demonstrates that our model captures local structure variations so well that it is comparable with geometric models. Moreover, geometric measurements are usually defined in very local regions. For instance, the helical-coordinate system is limited to one or two basepairs. However, our LWPH can quantitatively characterize structure information in local regions or domains with arbitrary sizes and shapes, where traditional geometrical measurements fail.
\end{abstract}


\newpage

{\setcounter{tocdepth}{5} \tableofcontents}

\newpage

\section{Introduction}

The great advancement in biological sciences and technologies has led to the accumulation of unprecedented gigantic amount of biomolecular data. Generally speaking, biological data can be classified into several categories, including genomics, transcriptomics, proteomics and metabolomics, which are deposited in several major databanks. A quick look at these databanks gives us a general perspective of the great amount of biological data that is available. Currently, in GenBank, there are more than 100 millions gene sequences, which is more than 1 billion bases. In protein data bank (PDB)\cite{Berman:2000}, there are about 150,000 three-dimensional biomolecular structures. The availability of the tremendous amount of the biological data has posed unprecedented opportunities for researchers from all areas. However, with great opportunities come great challenges. The high dimensionality, complexity, and variety of the biological data have rendered most powerful traditional methods and models useless. However, data analysis methods and models, including statistical learning, machine learning, data mining, manifold learning, graph/network models, topological data analysis (TDA), etc, have provided great promise in big data era and became more and more popular in bioinformatics and computational biology in the past one or two decades. Among these models, TDA has drawn special attention from mathematicians and computational scientists due to its unique characteristics. Unlike the general data analysis models, TDA studies topological invariants, which are global intrinsic structure properties. Roughly speaking, TDA can identify the ``shape of the data", thus it works as a powerful tool for simplification and dimensionality reduction. The key component of TDA is persistent homology (PH), which is developed from computational topology and algebraic topology. By assigning a geometric measurement to topological invariants, PH provides a bridge between geometry and topology. Recently, PH based machine learning models have delivered one of the best results in protein-ligand binding affinity prediction, partition coefficients, and mutation-induced folding energy variation \cite{cang:2017topologynet,cang:2017integration,nguyen:2017rigidity,cang:2017analysis,cang:2018representability,wu:2018quantitative} and won champions in several categories in the recent D3R Grand Challenges \cite{nguyen2018mathematical}, which is widely regarded as the most difficult challenge in drug design.

The great success of the persistent homology based machine learning models depends on the multiscale topological features obtained from the biomolecular structures \cite{Edelsbrunner:2002,Zomorodian:2005,Zomorodian:2008}. Unlike all previous geometric and topological models, which either focus on local structure information or study qualitative global properties, persistent homology embeds the geometric information on the topological invariants, thus provides the first quantitative topological measurements. With the proposed filtration process, a series of nested simplicial complexes, which encoded with structural topological information from different scales, are generated in PH. These simplicial complexes provides a multiscale topological profile of the structures. Topological features, such as individual components, holes, circles, and voids, can be evaluated from them. More importantly, some features persist while other die quickly during the filtration. The ``lifespans" or ``persisting times" provides a size measurement of the topological features\cite{Dey:2008,Dey:2013,Mischaikow:2013}. With its unique power in data simplification and structure representation, PH has already been applied to various fields, including shape recognition \cite{DiFabio:2011}, network structure \cite{Silva:2005,LeeH:2012,Horak:2009}, image analysis \cite{Carlsson:2008,Pachauri:2011,Singh:2008,Bendich:2010,Frosini:2013}, data analysis \cite{Carlsson:2009,Niyogi:2011,BeiWang:2011,Rieck:2012,XuLiu:2012}, chaotic dynamics verification \cite{Mischaikow:1999}, computer vision \cite{Singh:2008}, computational biology \cite{Kasson:2007,YaoY:2009, Gameiro:2013,KLXia:2014c, KLXia:2015a,BaoWang:2016a, KLXia:2015c,KLXia:2015b}, amorphous material structures \cite{hiraoka:2016hierarchical,saadatfar:2017pore}, etc. Many powerful softwares, including JavaPlex \cite{javaPlex}, Perseus  \cite{Perseus}, Dipha \cite{Dipha}, Dionysus \cite{Dionysus}, jHoles \cite{Binchi:2014jholes}, GUDHI \cite{gudhi:FilteredComplexes}, Ripser \cite{bauer2017ripser}, PHAT \cite{bauer2014phat}, DIPHA \cite{bauer2014distributed}, R-TDA package \cite{fasy:2014introduction}, etc, have been developed. The persistent times of topological features can be represented or visualized by several models, including persistent diagram (PD) \cite{Mischaikow:2013}, persistent barcode (PB) \cite{Ghrist:2008barcodes}, persistent landscape \cite{Bubenik:2007,bubenik:2015}, persistent image \cite{adams2017persistence}, etc. However, traditional persistent homology models can use only one filtration parameter, thus limits their applications in revealing some heterogeneous properties. To overcome this problem, several multidimensional filtration PH models have been proposed. These models can greatly boost the performance of traditional PH models. Another important approach is to design weighted persistent homology (WPH). The essential idea of WPH is to introduce weight information, which reflects certain physical, chemical or biological properties, into the simplicial complex generation or homological generator calculation. In this way, the topological features, obtained from WPH, will characterize more heterogeneous biomolecular properties.

Generally speaking, the weighted persistent homology can be characterized into three major categories, vertex-weighted\cite{edelsbrunner1992weighted,bell2017weighted,guibas2013witnessed,buchet2016efficient,KLXia:2015c,xia2015multiresolution}, edge-weighted\cite{petri2013topological,Binchi:2014jholes,KLXia:2015c,KLXia:2014persistent,cang:2017topologynet,cang:2018representability}, and simplex-weighted models\cite{dawson1990homology,ren2018weighted,wu2018weighted}. For vertex-weighted models, a weight value is defined on each vertex. Among these methods, the weighted alpha complex is the first model that has been used in biomolecular structure characterization\cite{edelsbrunner1992weighted}. By assigning a weight value to each atom, a modified distance function can be proposed and further used to generate the weighted Voronoi cell. The weighted alpha complex is a subset of weighted Delaunay complex, which is the duel of weighted Voronoi diagram. Other vertex-weighted models consider different types of weighted distance functions \cite{bell2017weighted,guibas2013witnessed,buchet2016efficient}. In a weighted Vietoris-Rips and $\check{C}ech$ complex model, a new distance function is proposed as a minimal value of the scaled Euclidean distances between the position to all atoms\cite{bell2017weighted}. The inverse of the distance function represents the union of balls centered at the atom, and naturally induces weighted $\check{C}ech$ complexex through nerve theorem\cite{bell2017weighted,guibas2013witnessed}. Weighted Vietoris-Rips complexes can be constructed by scaling the Euclidean distances between any two atoms by their weights\cite{bell2017weighted}. Similarly, $k$-distance functions are proposed and can be used to produce weighted
$\check{C}ech$ complexes\cite{guibas2013witnessed}. To characterize the multiscale properties of the biomolecules, a multiscale rigidity function is proposed\cite{KLXia:2015c,KLXia:2015d,xia2015multiresolution}. Each atom is associated with a weighted kernel function with a scale parameter. A rigidity function is defined as the summation of all the kernel functions and can be used to generate a series of nested Morse complexes in persistent homology. Unlike vertex-weighted models,  edge-weighted models usually specify unique weight values on edges \cite{petri2013topological,Binchi:2014jholes}. Using weight value as a filtration parameter, Vietoris-Rips or clique complex can be defined as a maximal simplicial complex, whose 1-skeleton has weight values larger (or smaller) than a certain filtration value\cite{buchet2016efficient}. For weighted clique rank homology model\cite{petri2013topological}, a network/graph, with a weight value on each edge, is considered. Clique complex can be defined on the subgraph composed of edges with weight larger than filtration value. A series of physics-aware models are proposed to characterize interactions within and between biomolecules\cite{KLXia:2015c,KLXia:2014persistent,cang:2017topologynet,cang:2017integration,nguyen:2017rigidity,cang:2017analysis,cang:2018representability,wu:2018quantitative, nguyen2018mathematical}. Various modified or generalized distance matrixes are used in these models.
Further, simplex-weighted persistent homology models, which based on weighted simplicial homology, are proposed\cite{dawson1990homology,ren2018weighted,wu2018weighted}. In these models, weight values are defined on simplexes in different dimensions. To ensure the consistence of the homology definition, weight values on different simplexes need to satisfy certain constraints or relations, so that a weighted boundary operation can be well-defined.

In this paper, we propose a localized weighted persistent homology LWPH. Our LWPH model is inspired by the recent great success of element specific persistent homology (ESPH) models as mentioned above. Unlike all previous weighted persistent homology models, which treat biomolecules as an inseparable system, ESPH decomposes the structure into a series of sub-structures made of certain type(s) of atoms. The subnetworks or subgraphs, especially those from protein-ligand complexes, have been proved to capture important biological properties, such as hydrophobic or hydrophilic interactions. In this way, ESPH models have delivered amazing results in biomolecular data analysis \cite{cang:2017topologynet,cang:2017integration,nguyen:2017rigidity,cang:2017analysis,cang:2018representability,wu:2018quantitative, nguyen2018mathematical}. Further, our LWPH models are different from traditional persistent local homology (PLH)\cite{bendich2007inferring,bendich2012local,ahmed2014local,bendich2015multi,fasy2016exploring,munkres2018elements}. The PLH studies the relative homology groups between a topological space and its subspace. It is usually used to assess the local structure of a special point within a topological space. In our LWPH, the biomolecular structures and configurations are decomposed into a series of local domains, that may overlap with each other. The general persistent homology or weighted persistent homology analysis is then applied on each of these local domains. In this way, DNA local structure, dynamics and functional properties can be embedded in our LWPH models.
Our model has been used in the analysis of DNAs. It has been found that our LWPH based features can be used to successfully discriminate A-, B-, and Z-types of DNA. More importantly, our LWPH based PCA model can identify two configurational states of a DNA system in ion liquid environment, which can only be revealed by the complicated helical coordinate representation. The great consistence with the helical-coordinate model demonstrates that our model captures the local structure variations so well that it is comparable with geometric models. Moreover, geometric measurements are usually defined in very local regions, for instance the helical-coordinate system is limited to one or two basepairs. However, our localized weighted homology can quantitatively characterize structure information in a much larger domain, where traditional geometrical measurements fail.

The paper is organized as follows. Section \ref{sec:method} is the summation of weighted persistent homology methods and models, and their applications in biomolecular systems. A detailed description of WPH models, including vertex-weighted, edge-weighted, and simplex-weighted models are presented in Section \ref{sec:WPH}. The localized weighted persistent homology (LWPH) model is discussed in Section \ref{sec:LWPH}. Section \ref{sec:results} is devoted for the application of LWPH in DNA systems. The paper ends with a conclusion.

\section{Methods}\label{sec:method}
In this section, we will provide a brief review of the weighted persistent homology models and their applications in biomolecular data analysis. After that, a detailed discussion of our localized weighted persistent homology model will be presented.

\subsection{Weighted persistent homology}\label{sec:WPH}
The essential idea of weighted persistent homology models is to introduce a specially-designed weight function/parameter that incorporates the biomolecular physical, chemical or biological properties, into the construction of simplicial complexes or homology generator evaluation. Generally speaking, all these WPH models can be classified into three types, including vertex-weighted\cite{edelsbrunner1992weighted,bell2017weighted,guibas2013witnessed,buchet2016efficient,KLXia:2015c,xia2015multiresolution}, edge-weighted\cite{petri2013topological,Binchi:2014jholes,KLXia:2015c,KLXia:2014persistent,cang:2017topologynet,cang:2018representability}, and simplex-weighted models\cite{dawson1990homology,ren2018weighted,wu2018weighted}.

To facilitate our discussion, we define a weighted point set as $(X,V)$ with $X=\{x_i|_{ i=1, 2, ..., N}\}$ and $V=\{v_i|_{ i=1, 2, ..., N}\}$. For each point $x_i$, a weight $v_i$ is assigned to it. We use $d(x_i,x_j)$ to represent the Euclidean distances between two points $x_i$ and $x_j$. Biologically, the weight $V$ is usually chosen to be the radius, atom number, etc.
\subsubsection{Vertex-weighted persistent homology}


\paragraph{Weighted alpha complex}
In weighted alpha complex\cite{edelsbrunner1992weighted}, a weighted distance is defined as $d^{\alpha}(x,x_i)=\sqrt{v_i^2+d(x,x_i)^2}$.
The weighted Voronoi region or Voronoi cell can be defined as
$$
VC_i=\{x | d^{\alpha}(x, x_i)\leq d^{\alpha}(x, x_j), for ~ all~ i\neq j \}.
$$
A $t$-weighted closed ball for $x_i$ is defined as
$\bar{B}^{\alpha}_i(t)=\{ x | d^{\alpha}(x,x_i) \leq t \}.$
The intersection of $t$-weighted closed balls and Voronoi cells is $R_{i}(t)=\bar{B}^{\alpha}_i(t) \bigcap VC_i$. In this way, the weighted alpha complex can be expressed as,
$$
A(t)= \{\sigma | \bigcap_{x_i \in X} R_{i}(t)\neq 0 \}.
$$
Essentially, the weighted alpha complex is a subsect of weighted Delaunay complex, which is the dual of weighted Voronoi diagram.

\paragraph{Weighted Vietoris-Rips and $\breve{C}$ech}
Bell, etc, have proposed a weighted Vietoris-Rips model and a weighted $\breve{C}$ech model\cite{bell2017weighted}. The weighted $\check{C}ech$ complex is defined as $\check{C}ech(X,V)=\mathcal{N}\{\bar{B}(x_i,v_i)|_{ i=1, 2, ..., N}\}$. Here $\bar{B}(x_i, v_i)=\{ x | d(x,x_i) \leq v_i \}$ is the closed ball centered at $x_i$ with radius $v_i$. The nerve $\mathcal{N}$ is the abstract simplicial complex from the closed balls. The weighted Vietoris-Rips Complex is defined as $VR(X,V)=\{ \sigma \subset X | d(x_i,x_j) \leq  v_i+ v_j, for~all~ x_i,x_j \in \sigma ~with~ x_i \neq x_j\}$.

For a filtration parameter $t \geq 0$, weighted $\check{C}ech$ complex at scale $t$ can be denoted as
\begin{eqnarray}\label{eq:cech}
\check{C}ech(X,V,t)=\mathcal{N}\{\bar{B}(x_i,t v_i); i=1,2,..,N\},
\end{eqnarray}
and the weight Vietoris-Rips Complex at scale $t$ can be expressed as
$$ VR(X,V,t)=\{ \sigma \subset X | d(x_i,x_j) \leq  t v_i+ t v_j, for~all~ x_i,x_j \in \sigma ~with~ x_i \neq x_j\}. $$

Moreover, we can define a distance function as $f_{X,V}(x)=\min_{x_i \in X}\{\frac{d(x,x_i)}{v_i}\}$. In this way, we have the inverse function
\begin{eqnarray}\label{eq:dist_function_v1}
f_{X,V}^{-1}([0,t])=\bigcup_{x_i \in X}\bar{B}(x_i,t v_i),
\end{eqnarray}
and it is homotopy equivalent to $\check{C}ech(X,V,t)$ as in Eq. (\ref{eq:cech}).

Computationally, the weighted Vietoris-Rips Complex\cite{bell2017weighted} can be constructed by using a weighted distance matrix $M=\{M_{ij}|_{i,j=1,2,...,N}\}$ with
$$ M_{i,j}=\frac{d(x_i,x_j)}{v_i+v_j}.$$
Various softwares, such as JavaPlex \cite{javaPlex}, Perseus  \cite{Perseus}, Dipha \cite{Dipha}, GUDHI \cite{gudhi:FilteredComplexes}, Ripser \cite{bauer2017ripser}, PHAT \cite{bauer2014phat}, DIPHA \cite{bauer2014distributed}, R-TDA package \cite{fasy:2014introduction}, can use distance matrix as their input data.

\paragraph{$k$-distance based model}
\begin{defn}
For any point set $X$ and $k$ is nonnegative integer, the $k$-distance\cite{guibas2013witnessed,buchet2016efficient} can be denoted as
$$ d_{X,k}^2(x)=\frac{1}{k} \sum_{x_i\in NN_{X}^k(x)} d^2(x, x_i) $$
with $NN_{X}^k(x)$ denotes the $k$ nearest neighbors in $X$ to the point $x$.
\end{defn}
Further, it can be expressed as power distance as follows,
$$ d_{X,k}^2(x)=\min\{ d^2(x,\bar{x})-w_{\bar{x}}; \bar{x} \in Bary^k(X)\}. $$
Here $Bary^k(X)$ denotes the barycenters of any subsets of $k$ points of $X$ and $w_{\bar{x}}=-\frac{1}{k}\sum_{1\leq i \leq k} d^2(\bar{x},x_i)$. Moreover, the sublevel sets of the $k$-distance $d_{X,k}$ are finite union of balls,
\begin{eqnarray}\label{eq:dist_function_v2}
d_{X,k}^{-1}([0,t])=\bigcup_{\bar{x} \in Bary^k(X)} B(\bar{x},(t^2+w_{\bar{x}})^{1/2}).
\end{eqnarray}
Similar to Eq. (\ref{eq:dist_function_v1}), the inverse of this distance function is homotopy equivalent to a weighted $\check{C}ech$ complex, which is the nerve of the closed balls\cite{guibas2013witnessed}.

\paragraph{Rigidity function based models}
Multiscale topological simplification models have been proposed\cite{KLXia:2015c,KLXia:2015d,xia2015multiresolution}. The key part of these models is multiscale rigidity function,
$$
\mu( x)=\sum_{i}^{N}v_i \Phi (d(x,x_i);\eta_i).
$$
Here $\eta$ is the scale parameter and $\Phi (d(x,x_i);\eta_i)$ can be chosen from any monotonically-decreasing functions, such as the generalized power-law equation,
\begin{eqnarray}\label{eq:couple_matrix25}
\Phi (d(x,x_i);\eta_i,\nu)=\frac{1}{1+(\frac{d(x,x_i)}{\eta_{i}})^\nu},
\end{eqnarray}
and the generalized exponential equation,
\begin{eqnarray}\label{eq:couple_matrix25}
\Phi (d(x,x_i);\eta_i;\nu)=e^{-(\frac{d(x,x_i)}{\eta_{i}})^\nu}.
\end{eqnarray}
Unlike previous distance functions in Eqs. (\ref{eq:dist_function_v1}) and (\ref{eq:dist_function_v2}), the inverse of rigidity function $\mu^{-1}([0,t])$ may not be expressed as a union of balls. However, it can generate Morse complexes. Computationally, the discrete Morse models can be used to evaluate the persistent homology.

\subsubsection{Edge-weighted persistent homology}
The essential idea for the edge-weighted persistent homology models is to assign a weight to each edge. With weight as  filtration parameter, the Vietoris-Rips complex can be defined as the maximal simplicial complex whose 1-skeleton has weight values larger (or smaller) than the filtration value. Computationally, a weighted distance matrix is usually proposed. The filtration is achieved through the increasing (or decreasing) of the weighted distance value.

\paragraph{Weighted clique rank homology}
The weighted clique rank homology is defined on weighted complex networks\cite{petri2013topological,Binchi:2014jholes}. For weighted networks, each edge/link has a weight on it. The filtration goes from the largest weight to the lowest one. At each filtration value $t$, a subgraph composed of edges with weight larger than $t$ is formed. Based on the subgraph, clique complex can be constructed. In this way, with the decrease of filtration value, a series of clique complexes are built and their homology and persistence can be calculated.


\paragraph{Physics-aware models}

Recently, a series of new persistent homology models have been proposed to characterize the various physical interactions within and between biomolecules\cite{KLXia:2015c,KLXia:2014persistent,cang:2017topologynet,cang:2018representability}. In these models, the distance matrix between atoms is modified based on their physical properties, including covalent bonds, protein-ligand interactions, electrostatic interactions, etc. To avoid confusion, we call them as physics-aware persistent homology.

For a biomolecule or biomolecular complex, we denote their atomic coordinates as $X=\{x_i|_{i=1,2,...,N}\}$, a distance matrix can be constructed as $M=\{M_{ij}=d(x_i,x_j)|_{i,j=1,2,...,N}\}$. Various modified distance matrices are proposed to characterize different physical, chemical and biological properties of the biomolecular structure.

\begin{defn} Multi-level persistent homology model considers a modified distance matrix as follows,
\begin{eqnarray}\label{eq:distance matrix_v1}
M_{ij}=\begin{cases} \begin{array}{ll}
             d(x_i,x_j),  & {\rm if ~ atoms ~} i~ {\rm and}~ j~ {\rm are~ not~bonded}; \\		
             {\infty}, & {\rm if ~ atoms ~} i~ {\rm and}~ j~ {\rm are~ bonded}.
	      \end{array}
\end{cases}
\end{eqnarray}
In computation, we can take ${\infty}$ as any value that is large than the filtration size.
\end{defn}

More generally, we can have a $n$-th level matrix as
\begin{eqnarray}\label{eq:distance matrix_v11}\nonumber
M_{ij}=\begin{cases} \begin{array}{ll}
             \infty,  & d(x_i,x_j)\leq n; \\		
             d(x_i,x_j),  & {\rm otherwise}.
	      \end{array}
\end{cases}
\end{eqnarray}

It has been found that when the modified matrices are employed, the barcode representation is significantly enriched and is able to capture the tiny structure perturbation between the conformations. Further, an interactive persistent homology model is proposed for protein-ligand binding analysis.

\begin{defn} An interactive persistent homology model is based on the revised distance matrix as follows,
\begin{eqnarray}\label{eq:distance matrix_v2}
M_{ij}=\begin{cases} \begin{array}{ll}
             d(x_i,x_j),  & {\rm if ~ atoms ~} i~ {\rm and}~ j~ {\rm ~from~different~molecules}; \\		
             \infty, & {\rm otherwise}.
	      \end{array}
\end{cases}
\end{eqnarray}
\end{defn}
In this way, the topological invariants induced by the interactions between two molecules, such as protein-protein, protein-DNA/RNA, protein-ligand, DNA/RNA-ligand, etc, can be well-captured.

All the function based persistent homology models in ESPH model\cite{cang:2018representability} can be generalized as $M_{ij}=\Phi(x_i,x_j)$. Here $\Phi(x_i,x_j)$ can be any function properties, including van der Waals interaction, electrostatic potential, or other generalized correlations.
%
%

\subsubsection{Simplex-weighted persistent homology}

\paragraph{Weighted simplicial homology}
Weighted simplicial homology is a generalization of simplicial homology\cite{dawson1990homology,ren2018weighted,wu2018weighted}. Every simplex has a weight in a ring $R$, and the boundary map is weighted accordingly. When all the simplices have the same weight $ a \in R \backslash \{0\}$, the resulting weighted homology is the same as the usual simplicial homology. We list some of the key definitions and results below.

\begin{defn}
A weighted simplicial complex (or WSC for short) is a pair $(K, w)$ consisting of a simplicial complex $K$ and a weight function $w : K \rightarrow R$, where $R$ is a commutative ring, such that for any $\sigma_1$, $\sigma_2$ with $\sigma_1 \subseteq \sigma_2$, we have
$w(\sigma_1) | w(\sigma_2)$.
\end{defn}

\begin{thm}
Let $I$ be an ideal of a commutative ring $R$. Let $(K,w)$ be a weighted simplicial complex, where $w : K \rightarrow R$ is a weight function. Then $K \backslash w^{-1}(I)$ is a simplicial subcomplex of $K$.
\end{thm}

For the definition of homology of weighted simplicial complexes, we require $R$ to be an integral domain with 1.

\begin{defn}
The weighted boundary map $\partial_n : C_n(K) \rightarrow  C_{n-1}(K)$ is the map:
\begin{eqnarray}\nonumber
\partial_n(\sigma)=\sum_{i=0}^{n}\frac{w(\sigma)}{w(d_i(\sigma))}(-1)^i d_i(\sigma)
\end{eqnarray}
where the face maps $d_i$ are defined as:
\begin{eqnarray}\nonumber
d_i(\sigma) = [v_0,...,\hat{v_i},..., v_n] ~({\rm deleting~ the~ vertex~ }v_i)
\end{eqnarray}
for any n-simplex $\sigma = [v_0,..., v_n]$.
\end{defn}

\begin{thm}
Let $f : K \rightarrow  L$ be a simplicial map. Then $f_\sharp \partial = \partial f_\sharp$,
where $\partial$ refers to the relevant weighted boundary map.
\end{thm}

\begin{defn}
 We define the weighted homology of a WSC to be
\begin{eqnarray}\nonumber
H_n(K,w)= ker(\partial_n)/ Im(\partial_{n+1}),
\end{eqnarray}
where $\partial_n$ is the weighted boundary map.
\end{defn}

\begin{prop}
Proposition. If all the simplices in $(K,w)$ have the same weight $a \in R \backslash \{0\}$, the weighted homology functor is the same as the usual simplicial homology functor.
\end{prop}

\paragraph{Weighted persistent Homology}
Given a weighted filtered complex $(K,w) =\{(K^i,w)\}_{i=0}$, for the $i$-th complex $K^i$, we have the associated weighted boundary maps $\partial^i_k$
and groups $C^i_k$, $Z^i_k$, $B^i_k$, $H^i_k$ for all integers $i$, $k = 0$.

\begin{defn} The weighted boundary map $\partial_i^k$, where $i$ denotes the filtration index, is the weighted boundary map of the ith complex $K^i$. That
is, $\partial^i_k$ is the map $\partial^i_k: C_k(K_i,w) \rightarrow C_{k-1}(K_i,w)$. The chain group $C_i^k$ is the group $C_k(K_i,w)$. The cycle group $Z^i_k$ is the group $ker(\partial_i^k)$, while the boundary group $B_i^k$ is the group $Im(\partial^i_{k+1})$. The homology group $H^i_k$ is the quotient group $Z^i_k/B^i_k$.
\end{defn}

\begin{defn} The p-persistent kth homology group of $(K,w) =\{(K_i,w)\}_{i=0}$ is defined as
\begin{eqnarray}\nonumber
H_k^{i,p}(K,w) := Z_k^i/(B_k^{i+p}\bigcap Z_k^i)
\end{eqnarray}
\end{defn}

\subsection{Localized weighted persistent homology}\label{sec:LWPH}
In all the above WPH models, weights are defined on the whole system to reveal the intrinsic global structural properties. Stated differently, all these models treat a biomolecule structure as an inseparable system, and explore their topological properties from the whole structure. In contrast,  ESPH models\cite{cang:2017topologynet,cang:2017integration,nguyen:2017rigidity,cang:2017analysis,cang:2018representability,wu:2018quantitative, nguyen2018mathematical} decompose a biomolecular structure into a series of sub-structures made of certain type(s) of atoms. It has been found that the generated subnetworks or subgraphs, especially those from protein-ligand complexes, can capture important biological properties, such as hydrophobic or hydrophilic interactions\cite{cang:2018representability}. Different from WPH and ESPH models, persistent local homology considers the relative homology groups between a topological space and its subspace. Usually, persistent local homology focuses on the local structure around a special point in a certain topological space.

Motivated by the success of ESPH models, we introduce our localized weighted persistent homology. Instead of decomposing biomolecular structures by their atom types, we focus on local biomolecular regions or domains and study their topological properties. For a better introduction of our LWPH, we will briefly review ESPH and PLH first.

\paragraph{Element specific persistent homology}
Biomolecules are made of various atoms with different properties. In protein, DNA or RNA, there are five common types of atoms, including $C$, $N$, $O$, $P$, and $S$, and several metal ions, such as $Fe$, $Mn$, $Zn$, etc. Ligands are small molecules, that interact with the protein, DNA or RNA. Other than the common five types of atoms, they may have some other unique atoms, such as $P$, $F$, $Cl$, $Br$, $I$, etc.

Generally speaking, ESPH is proposed to characterize the topological properties within the structure formed by one or several types of atoms\cite{cang:2017integration,cang:2017analysis,cang:2017topologynet,cang:2018representability,wu:2018quantitative}. Mathematically, it is achieved by assign weight value 1 to the selected types of atoms, and weight value 0 to all the rest atoms. For instance, all $C$ atoms from both protein and ligand can be selected to form $C$-networks or graphs. The topological properties of these structures characterize the hydrophobic interactions between protein and ligands. Similarly, hydrophilic interactions can be well captured by networks from protein nitrogen atoms and ligand oxygen atoms.

One of the most important properties for ESPH is to generate a series of sub-structures from the biomolecule, and systematically explore their topological properties. These element based sub-structures reveal more structure information that is directly related to the biomolecular physical, chemical, and biological properties.

\paragraph{Persistent local homology}
The persistent local homology \cite{bendich2007inferring,bendich2012local,ahmed2014local,bendich2015multi,fasy2016exploring} is based on the algebraic topological concept called local homology groups \cite{munkres2018elements}.

\begin{defn}
If $X$ is a space and if $x \in X$ is a point, then the local homology groups of $X$ at $x$ are the singular homology groups $H_k(X,X-x)$.
\end{defn}

From the excision theorem, we have the following Lemma.

\begin{lem}
Let $X$ be a Hausdorff space and let $A \subset X$. If $A$ contains a neighborhood of the point $x$, then $H_k(X,X-x)\simeq H_k(A,A-x)$. Therefore, for Hausdorff spaces $X$ and $Y$, if $x \in X$ and $y \in Y$ have neighbourhoods $U$, $V$, respectively, such that $(U, x)$ is homeomorphic to $(V,y)$, then the local homology groups of $X$ at $x$ and of $Y$ at $y$ are isomorphic.
\end{lem}

Generally speaking, local homology is used to evaluate the local structure of a topological space. Persistent local homology can be used in dimension reduction and manifold dimension detection.

\begin{figure}
\begin{center}
\begin{tabular}{c}
\includegraphics[width=0.7\textwidth]{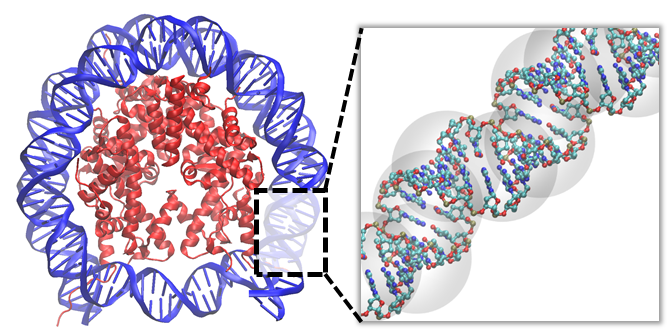}
\end{tabular}
\end{center}
\caption{The illustration of localized persistent homology. Instead of studying the topology for the whole structure, LPH focuses on the topology of each local region, i.e., the domain inside each transparent sphere. It should be noticed that these local regions may overlap with each other.
}
\label{fig:LPH}
\end{figure}

\paragraph{Localized weighted persistent homology}
The essential idea of our LWPH is to focus on local regions, of biomolecular structures or configurations, that incorporate the important physical, chemical or biological information, and perform the persistent homology analysis on them. In many situations, there may be several domains that are of great interest and we need to perform LPH on each domain. More generally, we can decompose biomolecular structures into a series of (overlapping) domains. For each domain, we carry out LWPH on all (or certain type(s) of) atoms that are of particular interest. In this way, a series of local topological properties can be obtained and we call them localized topological fingerprints. Figure \ref{fig:LPH} illustrates the idea of LWPH on a biomolecular complex. Each transparent sphere encloses in it a local biomolecular region. The persistent homology or other weighted persistent homology models as discussed above can then be employed on these local regions. To avoid confusion, if the general persistent homology is considered, we call it localized persistent homology (LPH). If a weighted persistent homology is considered, we call it localized weighted persistent homology (LWPH).

More specifically, for a biomolecule or biomolecular complex with atomic coordinates $X=\{x_i|_{i=1,2,...,N}\}$. The coordinate set $X$ can be decomposed into a series of domains $X^I$, with $X=\bigcup_{I=1}^{m} X^I$. Similar distance matrix $M^I$ as in Eqs (\ref{eq:distance matrix_v1}) and (\ref{eq:distance matrix_v2}) can be constructed on each of the domain. In this paper, we will focus on the study of DNA, which is made of paired nucleic acids. We can consider the weighted distance matrix on the domain $X^I$ as follows,
\begin{eqnarray}\label{eq:distance matrix_DNA}
M^I_{ij}=\begin{cases} \begin{array}{ll}
             d(x_i,x_j),  & x_i, x_j \in X^I, {\rm ~ atoms ~} i~ {\rm and}~ j~ {\rm ~are~from~different~nucleic~acid~{\color{red}residues}}; \\		
             \infty, & {\rm otherwise}.
	      \end{array}
\end{cases}
\end{eqnarray}
By using different weight values, LWPH can be designed to capture various local properties in biomolecular structures. The application of LPH and LWPH can be found in following sections.

It should be noticed that we still regard our LPH as a weighted persistent homology model. Essentially, we can set weight as a vector made of zeros and ones. Each vector has $m$-component, which $m$ the number of local regions. Each atom has a weight vector that assigns it to certain local region or regions. In this way, we can focus on certain domains of a biomolecule, or decompose the entire biomolecule into a series of local regions.

\section{Results and discussions}\label{sec:results}
In this section, we discuss the application of our localized persistent homology and localized weighted persistent homology in the study of DNA structures. To avoid confusion, only the weight definition as in Eq (\ref{eq:distance matrix_DNA}) is used in our LWPH. The persistent barcodes are used for the representation and visualization of LPH and LWPH results. The persistent Betti numbers (PBNs) are evaluated from barcodes and a systematical evaluation of PBNs under helical coordinates demonstrate the incorporation of the geometric information in LPH and LWPH. Further, we show that PCA of the feature vectors from the LWPH based PBNs can be successfully used in the classification of the A-, B-, and Z-types of DNAs. Moreover, we explore the DNA structure variations in both water and ion liquid (IL) environment using molecular dynamics. With LWPH based feature vectors, we can not only reveal the confinement effect of DNA configurations from water to IL environment, but also identify two DNA configurational states in IL environment. Detailed analysis found that all global-scale PCA models, including atom-coordinate based PCA, persistent homology based PCA, and element-specific persistent homology based PCA, all fail in clustering the DNA configurational states in IL. In contrast, LWPH based all-atom or selected-atom models are always able to characterize the DNA structure variations. The LWPH results are highly consistent with the helical-coordinate based PCA model.

\subsection{DNA local topological features}
\begin{figure}
\begin{center}
\begin{tabular}{c}
\includegraphics[width=0.7\textwidth]{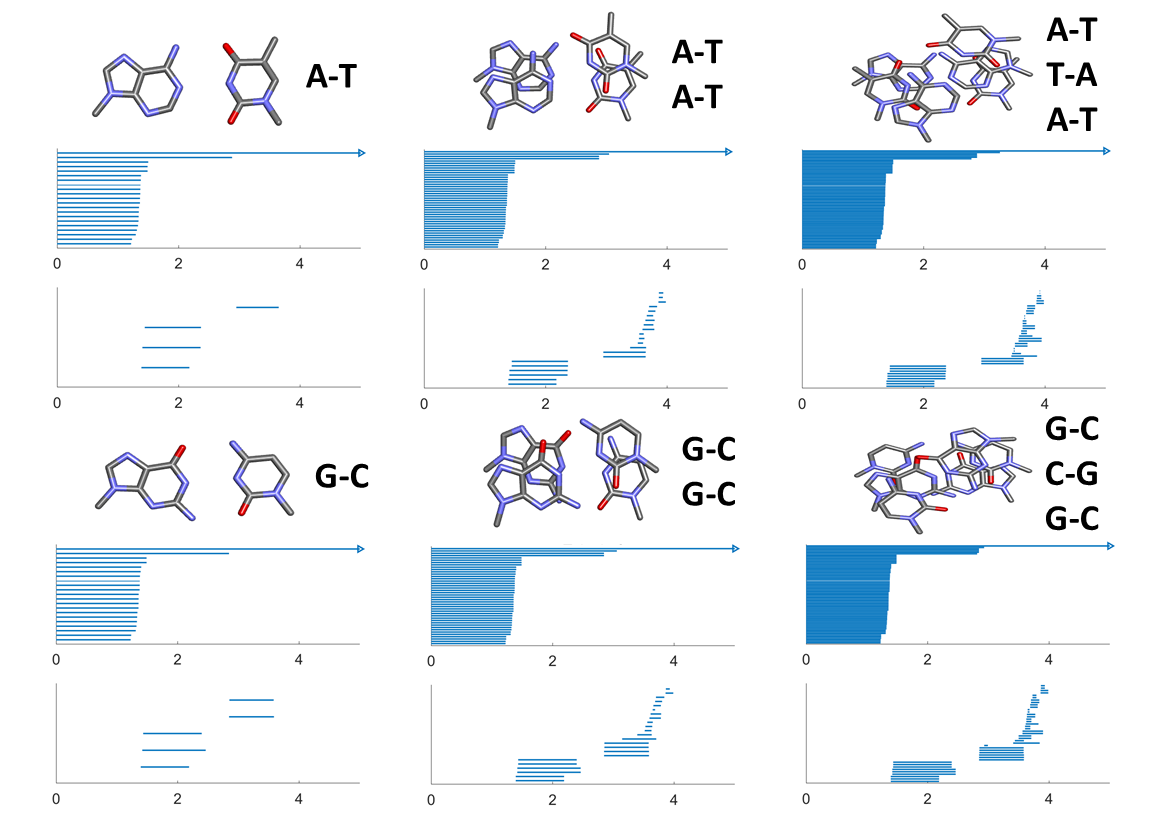}
\end{tabular}
\end{center}
\caption{The LPH based barcodes for different combination of DNA base pairs. It can be seen that A-T and G-C base pairs all have three local $\beta_1$ bars from around 1.4 \AA~ to 2.4 \AA. But they differ greatly in the global region, where A-T pair contributes one significant $\beta_1$ bar from around 2.9 \AA~ to 3.6 \AA,  while G-C pair generates two. These barcode fingerprints characterize the intrinsic DNA structure properties, i.e., local and global loop/ring motifs.
}
\label{fig:base_pairs}
\end{figure}

\begin{figure}
\begin{center}
\begin{tabular}{c}
\includegraphics[width=0.7\textwidth]{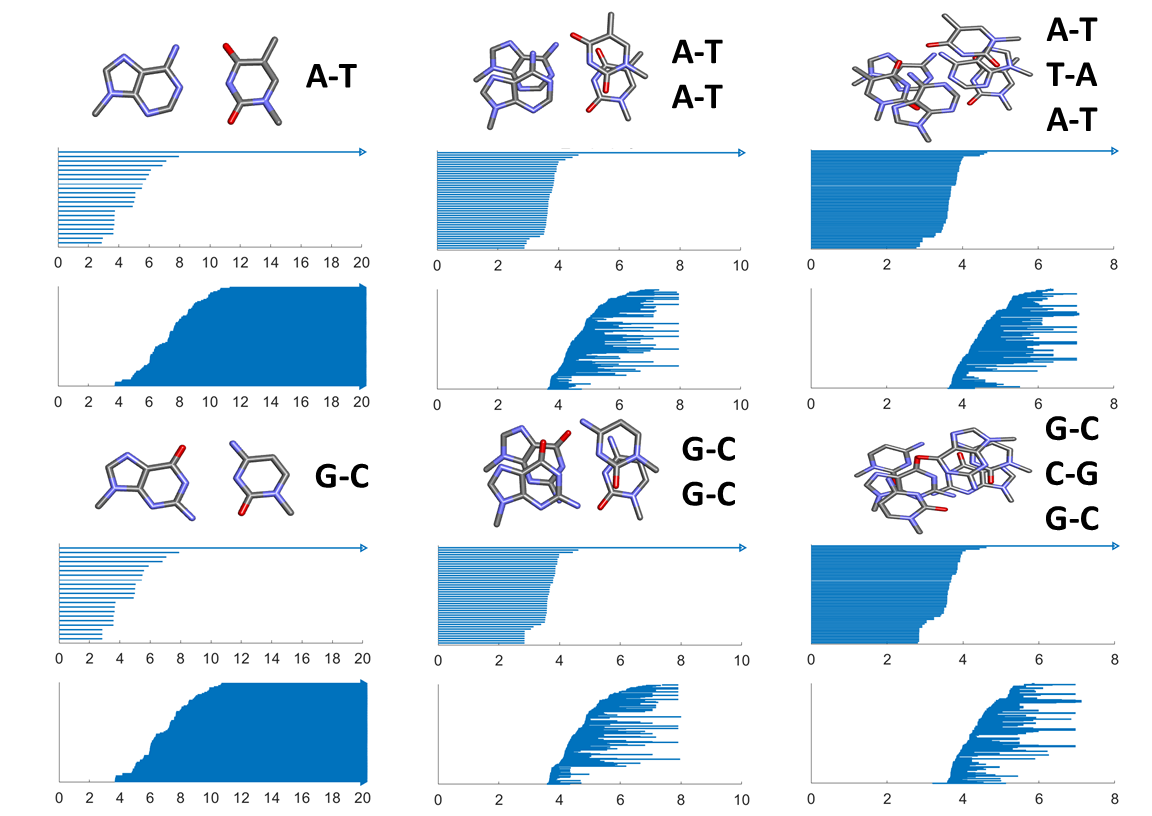}
\end{tabular}
\end{center}
\caption{The LWPH based barcodes for different combination of DNA base pairs. The weighted distance matrix in (\ref{eq:distance matrix_DNA}) is considered. Only the distances between two atoms from different nucleic acids are set to be their Euclidean distance. Others are zero. The shortest $\beta_0$ bars are distances between adjacent atoms from two bases, thus they characterize the hydrogen bonds between two nucleic bases. For a single base pair (A-T or G-C) situation, the generated $\beta_1$ bars will never be ``killed" as no 2-simplexes can be formed.
}
\label{base_pairs_newM}
\end{figure}

DNA molecule has a double helical structure composed of four types of nucleic acids, i.e., A, T, G, and C. It has remarkably different scales, ranging from nucleic bases, minor and major grooves, to larger structures like nucleosome, chromatin, and then to chromosome. We have proposed a multi-resolution PH model to characterize structural topological features or topological fingerprints of DNA structures in different scales\cite{KLXia:2015d,xia2015multiresolution}. However, the model focuses on the global DNA topology. In this section, we explore DNA topological features from local structures, i.e., DNA localized topological fingerprints. We study the barcodes for both LPH and LWPH models, i.e., one with common homology and the other with the weighted persistent homology as in Eq. (\ref{eq:distance matrix_DNA}). Different local structures can be systematically studied. In current paper, our focus is local topological features from different DNA base pairs or base steps. 

To facilitate a better description, we introduce some basic notations. In general, results from PH can be represented as pairs of ``birth" and ``death" times, i.e., the filtration values for homology generators to appear and disappear. We denote them as follows,
$$
L_{k}=\{ l^k_j|l^k_j=b^k_j-a^k_j; k \in \mathbb{N}; j \in \{1, 2, ..., N_k\} \},
$$
here $a^k_j$, $b^k_j$, $l^k_j$ to represent ``birth", ``death", and ``persistence" for $j$-th generator of $k$-th dimensional Betti number, respectively. And $N_k$ is the total number of $k$-th dimensional topological generators. Due to the limited atoms in a local structure, we only consider dimension $k$ equals to 0 and 1. As stated in introduction, results from PH models are usually visualized through persistent barcodes or persistent diagram. Further, different PH based functions are proposed for the visualization, representation and modeling of topological information\cite{Carlsson:2009,bubenik:2015,Chintakunta:2015}. The persistent Betti number (PBN) or Betti curve is one of them. It is defined as the summation of all the $k$-th dimensional barcodes,
\begin{eqnarray}\label{eq:PBN}
f_{\rm PBN}(x;L_{k})= \sum_{j} \chi_{[a^k_j,b^k_j]}(x)
\end{eqnarray}
Function $\chi_{[a^k_j, b^k_j]}(x)$ is a step function, which equals to one in the region $[a^k_j, b^k_j]$ and zero otherwise.

As a simple one-dimensional function, PBN has been used in data analysis for dimensionality and complexity reduction. Moreover, PBN based feature vectors can be input into various machine learning models\cite{cang:2017topologynet,cang:2017integration,nguyen:2017rigidity,cang:2017analysis,cang:2018representability,wu:2018quantitative, nguyen2018mathematical}. In this section, PBN based PCA models are used in DNA structure classification and trajectory clustering.

\begin{figure}
\begin{center}
\begin{tabular}{c}
\includegraphics[width=0.8\textwidth]{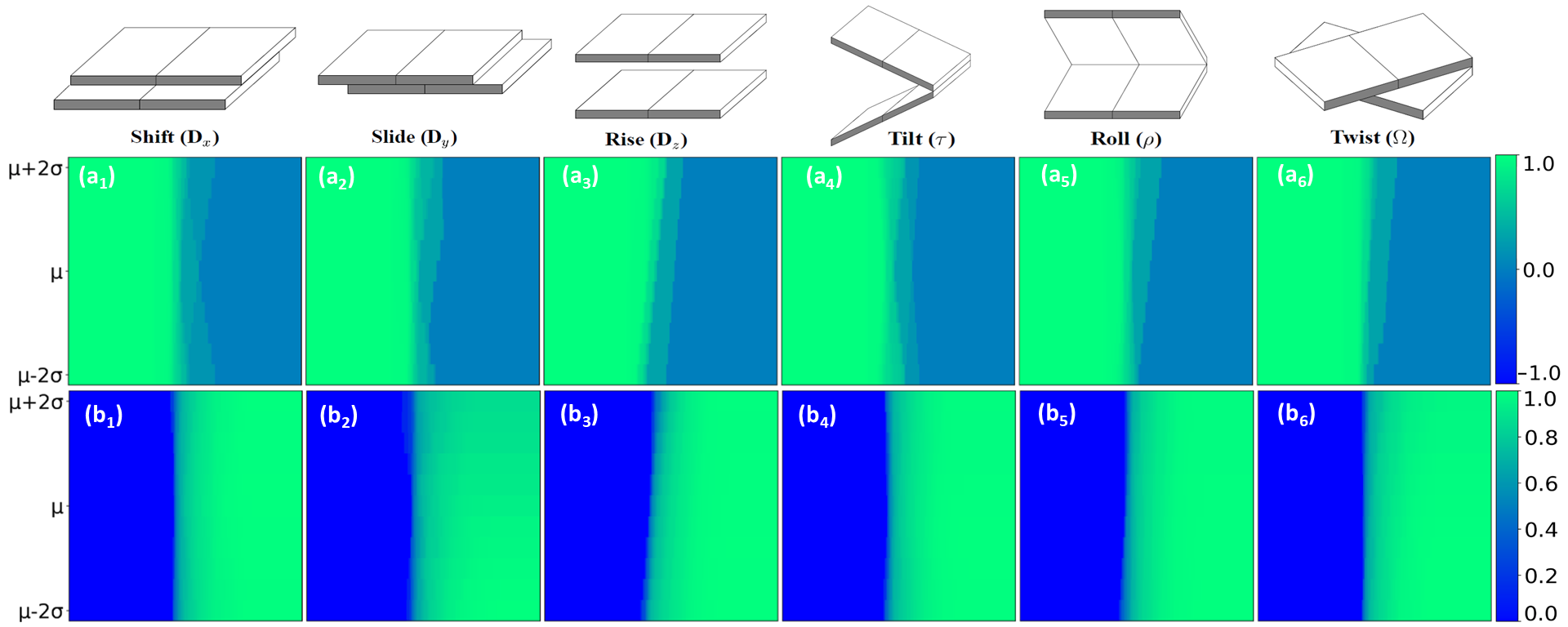}
\end{tabular}
\end{center}
\caption{The LPH based PBN image representation for each two-base-step helical parameter. In the $i$-th PBN image, we systematically change the $i$-th helical parameter value from  \(  \mu _{i}-2 \sigma _{i} \)  to  \(  \mu _{i}+2 \sigma _{i} \), with all other helical parameter remain as constant, to deliver a series of DNA structures. PBN can be calculated for each DNA structure and all of them stacked together to form a two-dimensional image. It can be seen that, both $\beta_0$ and $\beta_1$ PBN functions vary with the change of helical parameter value. And the change of $\beta_0$ PBN functions seem to be more dramatic.
}
\label{fig:helical_twoPair_orgM}
\end{figure}

\begin{figure}
\begin{center}
\begin{tabular}{c}
\includegraphics[width=0.8\textwidth]{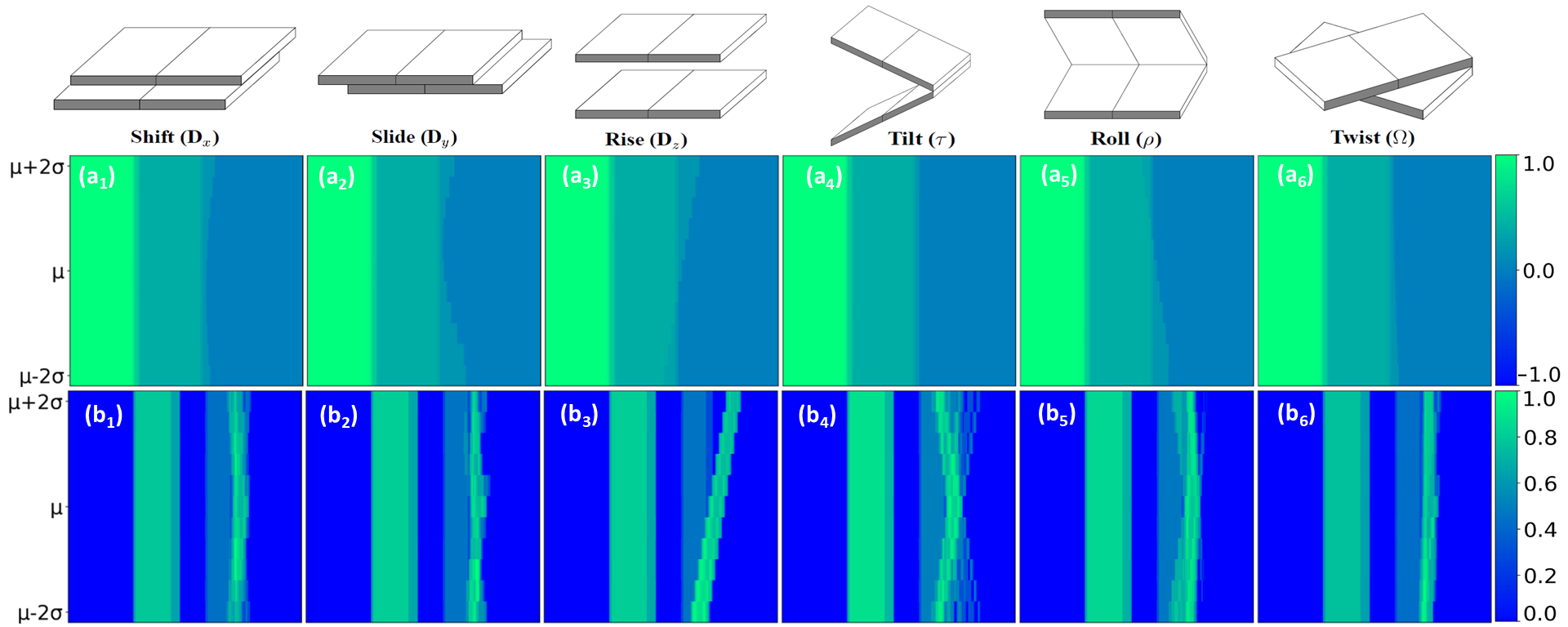}
\end{tabular}
\end{center}
\caption{The LWPH based PBN image representation for each two-base-step helical parameter. DNA structures are prepared in the same way as in Figure \ref{fig:helical_twoPair_orgM}. It can be seen that, similar to LPH based PBNs, LWPH based $\beta_0$ and $\beta_1$ PBN functions vary with the change of helical parameter value. However, the change of $\beta_1$ PBN functions seem to be more dramatic in LWPH models.
}
\label{fig:helical_twoPair_newM}
\end{figure}

\paragraph{LPH and LWPH for DNA analysis}
As stated above, we consider two types of LPHs, one is LPH with Euclidean distance matrix and the other is LWPH with a similar weighted distance matrix as in Eq. (\ref{eq:distance matrix_DNA}). For LWPH, we assume the distance between two atoms from same nucleic acid residue to be 0, and the distance between two atoms from different nucleic acid residues to be their Euclidean distance. In this way, our LPH characterizes more about covalent bond information, while our LWPH reveals more non-covalent bond properties between different nucleic acid residues. The DNA base pair structures are generated by using the 3DNA software\cite{lu20033dna}.

Figure \ref{fig:base_pairs} illustrates LPH barcodes for different combinations of base pairs. Shorter $\beta_0$ bars (with length around 1.0 \AA~to 1.5 \AA) correspond to covalent bonds, longer $\beta_0$ bars with length around 2.8 \AA~ correspond to the hydrogen bonds between paired bases. For $\beta_1$ bars, longer ones appear in earlier stage of filtration, i.e., from around 1.4 \AA~ to 2.4 \AA, correspond to sugar rings and nitrogenous base rings. The longer $\beta_1$ bars range from around 2.9 \AA~ to 3.6 \AA, representing loops between paired bases. It can be seen that each A-T pair only contributes one such longer $\beta_1$ bar, while each G-C pair contributes two.

Figure \ref{base_pairs_newM} illustrates the corresponding LWPH barcodes. Similar to LPH results, the total number of $\beta_0$ bars is exactly the number of atoms, and shorter $\beta_0$ bars with length around 2.8 \AA~ also correspond to hydrogen bonds between paired bases. Different from LPH results, more hydrogen-bond related $\beta_0$ bars appear in LWPH barcode than LPH barcodes. And the lengths of $\beta_0$ bars for LWPH are systematically longer than those for LPH model, indicating that more long-range interactions related information are preserved in our LWPH model. Moreover, the $\beta_1$ barcodes for LWPH are much more complicated. Their geometric meanings are not as straightforward as LPH models. Generally speaking, the $\beta_1$ bar in LWPH represents loop or ring structure with edges between different nucleic acids. In this way, a much larger amount of $\beta_1$ bars are generated. Moreover, when there are only two nucleic acids (or one base pair), the $\beta_1$ bars persist forever.

In general, LPH and LWPH characterize different structure properties, the former is more about covalent-bond related topology while the latter is more about topology from non-covalent bonds. Physically, covalent-bonds are much stronger than non-covalent-bond. In this way, LPH are relatively more ``stable" and less sensitive to structure variations under thermal fluctuations. While LWPH are less ``stable" and much easy to change if there is some external perturbations.

\paragraph{LPH and LWPH based DNA representation}

With the embedded geometric information, LPH and LWPH can be used in not only qualitative but also quantitative representation of different structures. To assess LPH and LWPH based quantitative representation, we systematically generate a series of DNA base-pair configurations using DNA helical coordinates. According to CEHS scheme\cite{lu1997structure}, the motion of a base pair or two neighbouring base pairs can be depicted by 12 helical parameters, including 6 one-base-step related parameters, i.e., shear, stretch, stagger, buckle, propeller and opening, and another 6 two-base-step related parameters, i.e., shift, slide, rise, tilt, roll and twist. For each parameter, we prepare 11 DNA structures, with parameter value taken equally from  \(  \mu _{i}-2 \sigma _{i} \)  to  \(  \mu _{i}+2 \sigma _{i} \),  using 3DNA\cite{lu20033dna}. Here  \( \mu _{i} \) ,  \( \sigma _{i} \)  are the mean value and standard deviation of parameter \textit{i}. And the rest helical parameters remain as constants, i.e., their mean values. The mean value and standard deviation of each parameter can be obtained from crystal structure\cite{lu1997structure}. In DNA helical coordinate evaluation, only the base atoms and $C1'$ of the sugar ring are considered. For a fair comparison, the same atoms are used in our LPH and LWPH models.

We apply our LPH and LWPH on these series of DNA local structures and check if the quantitative structure variations are reflected in their barcodes. To facilitate a systematical comparison, we consider the PBN function in Eq.(\ref{eq:PBN}). For each helical parameter, we take the natural logarithm of all its PBN functions and stack them together to form a two-dimensional image. Figures \ref{fig:helical_twoPair_orgM} and \ref{fig:helical_twoPair_newM} illustrate the results of our LPH and LWPH for two-base-step parameters, respectively. It can be seen that instead of remaining unchanged for all helical parameter values, both $\beta_0$ and $\beta_1$ PBN functions for LPH and LWPH models vary greatly. This demonstrates that both models are sensitive to these structure differences. More specifically, in LPH based PBN, $\beta_0$ functions seem to have comparably larger variations than $\beta_1$ functions. In LWPH based PBN, $\beta_1$ functions seem to have greater variations than $\beta_0$ functions. We also check the $\beta_0$ and $\beta_1$ PBN functions for LPH and LWPH models for one base step. The results are demonstrated in Figures \ref{fig:helical_onePair_orgM} and \ref{fig:helical_onePair_newM} in Supplementary. Again, both functions in two models show variations with the change of helical parameter value.

Since the PBN functions from LPH and LWPH are sensitive to the structure variations, we can use them as measurements for DNA structure, function and dynamics analysis. In the following sections, PBN and PBN based features are used in the classification of DNA types and clustering of DNA trajectories.

\subsection{Local topological feature based DNA classification and clustering}
\begin{figure}
\begin{center}
\begin{tabular}{c}
\includegraphics[width=0.95\textwidth]{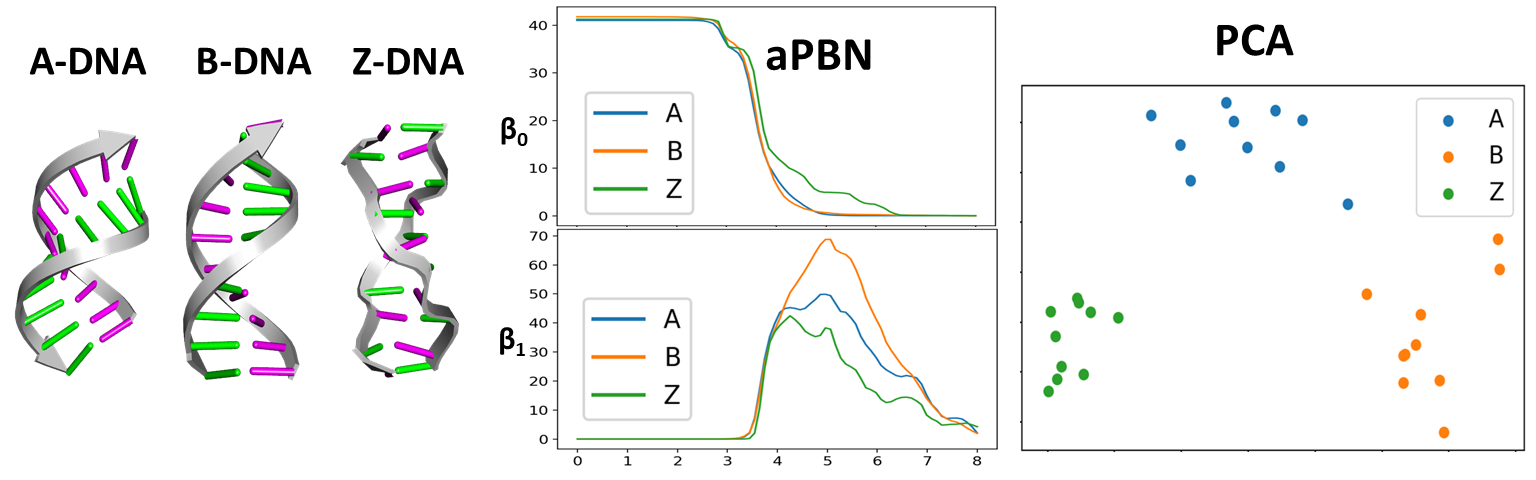}
\end{tabular}
\end{center}
\caption{LWPH based classification of three DNA types, i.e., A-DNA, B-DNA, and Z-DNA. The average persistent Betti number (aPBN) from our LWPH for three types of DNAs. We discretize the aPBN equally into a series of numbers and use these values as features for PCA. It can be seen that LWPH based aPBN and PCA results can clearly discriminate three DNA types. }
\label{fig:ABZ_LWPH}
\end{figure}

\subsubsection{Classification of three typical DNA forms}

We consider three types of DNA structures, including A-, B- and Z-forms. We randomly pick 10 PDB files for each form of DNA and the PDB IDs are shown in Table \ref{tab:PDBs} in the supplementary. In LPH and LWPH, the same atom combination of each base step is chosen as in the above case, i.e., base atoms and $C1'$ of the sugar ring.  The PBN function is calculated for each base step. To systematically compare the PBNs for three types of DNA forms, we summarize all the PBNs from the same DNA form and then take the average.

The results from LWPH are demonstrated in Figure \ref{fig:ABZ_LWPH}. It can be seen that the three PBN profiles have very different $\beta_1$ , particularly on the filtration range from 4.0 \AA~ to 7.0 \AA. Further, we consider the principal component analysis (PCA) for DNA classification. For each DNA structure, we define a vector made of the average $\beta_0$ and $\beta_1$ PBN values equally taken from 2.0 \AA~to 8.0 \AA~ with an interval 0.1\AA. In this way, a feature vector with 120 elements is defined for all 30 DNA structures. The PCA results are demonstrated in Figure \ref{fig:ABZ_LWPH}. Here x-axis and y-axis represent the first and second eigenvectors, respectively. It can be seen that three forms of DNA locate in the different regions with clear boundary, which further confirms that LWPH based features can distinguish the subtle conformational deviation. Figure \ref{fig:ABZ_LPH} in the supplementary shows the results from LPH. In comparison with LWPH, LPH based PBNs show no obvious difference, and PBN based PCA does not classify the dataset into three individual clusters.

\subsubsection{Clustering of DNA conformations in different environments}

\begin{figure}
\begin{center}
\begin{tabular}{c}
\includegraphics[width=0.9\textwidth]{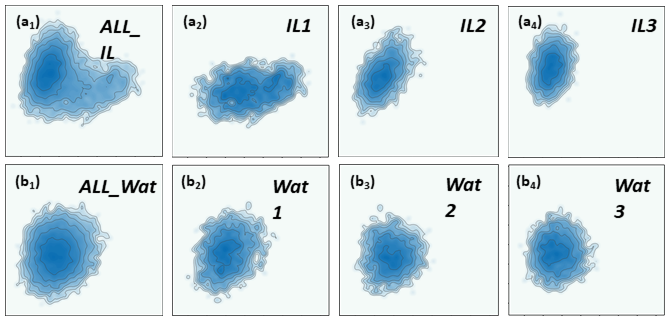}
\end{tabular}
\end{center}
\caption{The contour map generated from our LWPH based PCA models for DNA configurations in IL and WAT environments. The x-axis and y-axis are the first and second principal components. (${\bf a}_1$) The DNA configuration ensemble for all three trajectories for IL. (${\bf a}_2$) to (${\bf a}_4$) Three DNA configuration trajectories from the MD simulation with IL. (${\bf b}_1$) The DNA configuration ensemble for all three trajectories in water environement. (${\bf b}_2$) to (${\bf b}_4$) Three DNA configuration trajectories from the MD simulation using water solution. It can be seen clearly that areas of contour map in IL are much smaller than in water, indicating the confinement effect of the DNA configurations in IL. Further, contour graph for IL1 (${\bf a}_2$) shows a clear difference of that for IL2 (${\bf a}_3$) and IL3 (${\bf a}_4$), meaning there is a subtle change of the ion-DNA binding mode in trajectory IL1.}
\label{fig:PCA_LWPH}
\end{figure}

We have demonstrated the DNA structure classification with our LWPH. However, A-, B- and Z- forms of DNAs are static and have relatively ``large" configurational differences. In the following, we consider a more challenging case. That is the clustering of the molecular dynamics simulations of the same DNA molecule in different solvent environments. 

\paragraph{Molecular dynamics setting}
A brief introduction of the MD procedure is presented as follows. The initial structure of 16-mer DNA duplex is prepared using 3DNA\cite{lu20033dna} and centered in a cubic box. Two different solution environments are used, including ion liquid (IL) and water (WAT). For IL environment, 600 BMIM$^{+}$ and 600 BF$_4^-$ are firstly inserted and the box is then solvated with TIP3P water and Na$^+$. For WAT environment, the box is directly solvated with water and and Na$^+$. After the 100 ps thermostat and 100 ps barostat, the system then goes through a 100 ns product MD. Under each environment setting, we conduct 3 repeated MD simulations, so we obtain 6 trajectories in total. We denoted them as IL1-3 (trajectory 1 to 3 in IL) and WAT1-3 (trajectory 1 to 3 in water). All the simulations are conducted using GROMACS 4.6 package\cite{hess2008gromacs}. In our data analysis, 5000 sample frames evenly extracted from the last 10 ns trajectory of 6 simulations are considered. The detailed MD simulation setting and parameters can be found in the related paper\cite{meng2018molecular}.

\paragraph{Weighted persistent homology modeling}
For DNA conformation clustering, we extract 13 non-terminal DNA base steps for each frame of the simulation data. Similar to DNA-type classification case, for each base step, we construct a 120-element PBN feature from the LWPH. Then, we concatenate all 13 sets of PBN values together into feature vector for each DNA configuration. This LWPH based feature vector is used in the PCA of DNA trajectories in different environments. More specifically, a covariance matrix of feature vectors for all the frames is built up, and further eigen-decomposed into principal components (eigenvectors). The first two eigenvectors construct a plane and all feature vectors are projected to it. For comparison, in both IL and water, we apply PCA not only on three individual MD trajectories, but also the ensemble made of all three trajectories together. The results are demonstrated in Figure \ref{fig:PCA_LWPH}. The projected points are illustrated as their contour values for a better visualization. To avoid confusion, Figures \ref{fig:PCA_LWPH}(${\bf a}_1$) to (${\bf a}_4$) are for DNA trajectories in IL environment. Among them, Figure \ref{fig:PCA_LWPH}(${\bf a}_1$) is for the ensemble of all three trajectories. Figures \ref{fig:PCA_LWPH}(${\bf a}_2$) to (${\bf a}_4$) are for the three trajectories, respectively. Figures \ref{fig:PCA_LWPH}(${\bf b}_1$) to (${\bf b}_4$) are for DNA trajectories in water environment. Again Figures \ref{fig:PCA_LWPH}(${\bf b}_2$) to (${\bf b}_4$) are for the three trajectories, respectively, and Figure \ref{fig:PCA_LWPH}(${\bf b}_1$) is for the ensemble of all three trajectories.

\begin{figure}
\begin{center}
\begin{tabular}{c}
\includegraphics[width=0.9\textwidth]{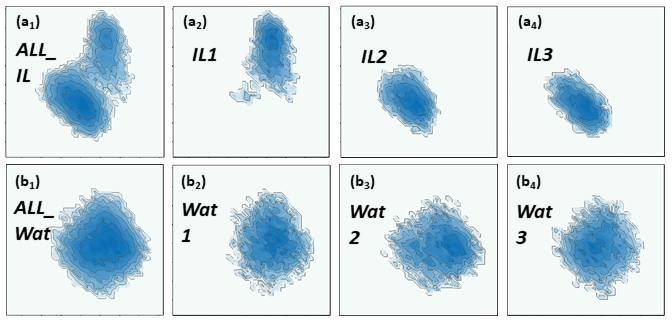}
\end{tabular}
\end{center}
\caption{The contour map generated from helical-parameter based PCA models for DNA configurations in IL and WAT environments. The x-axis and y-axis are the first and second principal components. (${\bf a}_1$) The DNA configuration ensemble for all three trajectories for IL. (${\bf a}_2$) to (${\bf a}_4$) Three DNA configuration trajectories from the MD simulation with IL. (${\bf b}_1$) The DNA configuration ensemble for all three trajectories in water environment. (${\bf b}_2$) to (${\bf b}_4$) Three DNA configuration trajectories from the MD simulation using water solution. The same confinement effect and two center distribution of DNA configurations in IL environment as in Figure \ref{fig:PCA_LWPH} are observed.}
\label{fig:PCA_helical}
\end{figure}

\begin{table}
  \centering
	\caption{The confinement effect in ion liquid solution. The area of distribution of each MD trajectory repeat. The area is counted as the number of grids with population larger than 5\% of the largest population grid. Four different methods are considered, including the atom coordinate based PCA (AC-PCA), the helical-parameter based PCA (HP-PCA), the localized persistent homology based PCA (LPH-PCA), and localized weighted persistent homology based PCA (LWPH-PCA). We have not listed the area for IL1 from HP-PCA and LWPH-PCA, as dynamic transition is observed in these representations and the area cannot be treated as the measurement of degree of conformational fluctuation.}
\begin{tabular}{|c|c|c|c|c|c|c|}
\hline
    &IL1 &IL2 &IL3  &Wat1 &Wat2 &Wat3 \\ \hline
AC-PCA	 &212&167&173&307&301&268\\ \hline
HP-PCA     &-& 217& 203& 350& 352&341\\ \hline
LPH-PCA          &340	&321 &333 &345 &357 &354 \\ \hline
LWPH-PCA         &-& 296& 295&554 &514 &469\\ \hline
    \end{tabular}
    \label{tab:Task1_BF_PCA_SVM}
\end{table}

\paragraph{Basic results}
Several unique properties can be seen from Figure \ref{fig:PCA_LWPH}. Firstly,the confinement effect, i.e., the reduction of distribution area, of IL can be clearly observed. In fact, we can count the area of the distribution in IL and water and the results are listed in Table \ref{tab:Task1_BF_PCA_SVM}. It can be seen that, the areas of distribution map in IL are up to 2/3 of that in water, the smaller area confirms the fluctuation of DNA in IL is much attenuated. Secondly, contour graphs for IL and WAT show significantly different patterns. For IL solution, two centers can be clearly identified from contour graph in Figure \ref{fig:PCA_LWPH}(${\bf a}_1$). Among them, one locates on upper left and the other locates on right part. In contrast, for water solution, only one center can be found and its position differs greatly from the ones of IL systems. The huge difference indicates the change of DNA conformations in different environments. Further, for water solution, all contour graphs have nearly the same distribution, indicating that all three trajectories behaved quite similarly. In contrast, for IL solution, contour graph for IL1 shows a clear difference of those for IL2 and IL3, meaning there is a subtle change of ion-DNA binding mode in trajectory IL1. Our LWPH based PCA is sensitive enough to capture this subtle structure variation.

Our LWPH based results are highly consistent with previous results from the helical parameter based model\cite{meng2018molecular}. As demonstrated in Figure \ref{fig:PCA_helical}, confinement effect is also observed in helical-parameter based contour graphs for IL solution. Further, helical-parameter results show two different centers, while WAT contour graphs have only one center. This means that, helical parameter model also captures the subtle change of ion-DNA binding mode in trajectory IL1. To further check if our LWPH and helical-parameter based models identify the same type of ion-DNA configurational changes, we decompose the contour graphs of IL1 into 10 separated subgraphs, each of them represent the DNA trajectories in 1 ns. We can clearly identify four center regions from these subgraphs, and they are consistent with results from helical-parameter based PCA. This further indicates that our LWPH based models are highly sensitive to DNA local structural variations. The results are demonstrated in Figures \ref{fig:MD_region_LWPH} and \ref{fig:MD_region_LWPH_CG} in the supplementary.

\begin{figure}
\begin{center}
\begin{tabular}{c}
\includegraphics[width=0.5\textwidth]{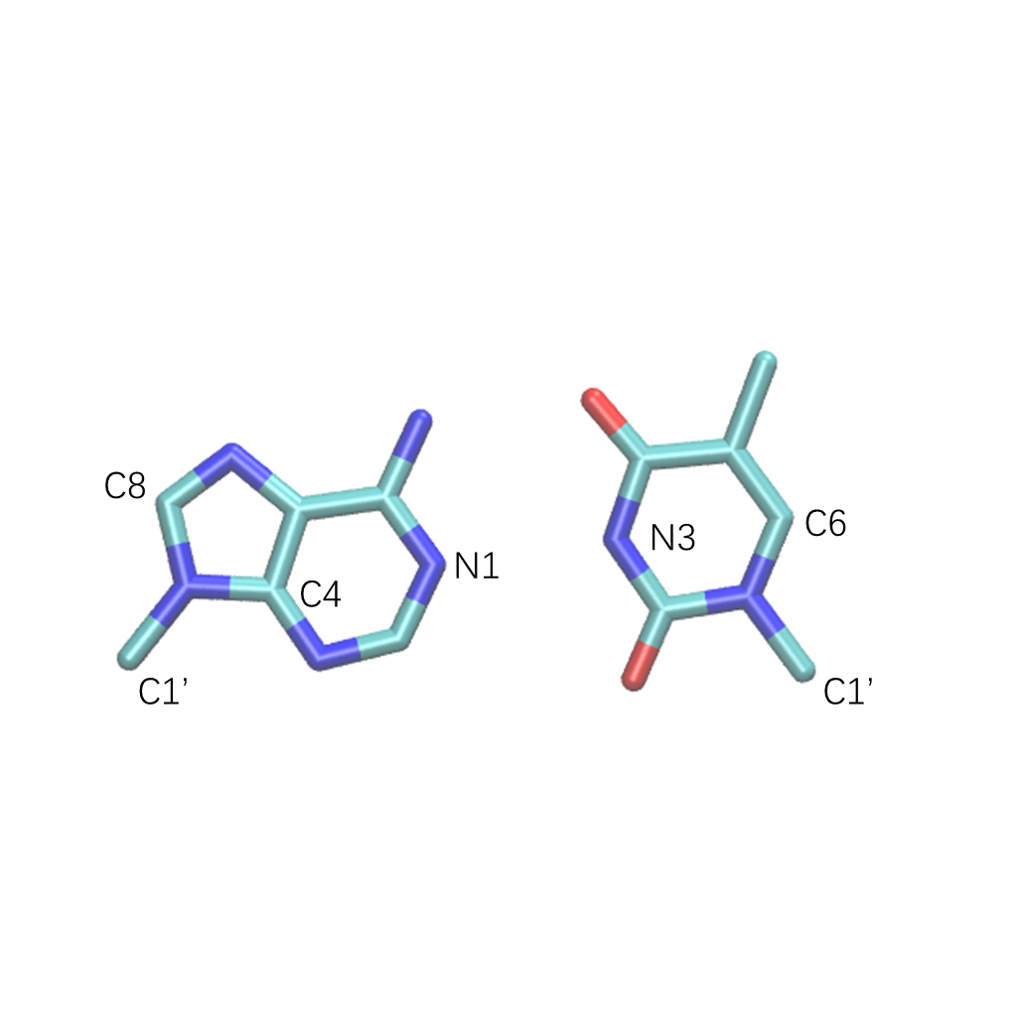}
\end{tabular}
\end{center}
\caption{The atoms used for selected atom based LWPH computation, the selected atoms are marked. }
\label{fig:base_pick}
\end{figure}

Further, it should be noticed that the local DNA structure variations in IL1 cannot be captured by the general global models. As demonstrated in Figure \ref{fig:PCA_all_atom} in the supplementary, the general atom-coordinates based PCA fails to capture the DNA structure variation in IL1. Similarly, LPH based models fail to reveal the variation. Moreover, it even cannot reveal the confinement effect of IL environment.  This is largely due to the reason that LPH based model focuses more on the covalent bonds and its related structures. Even though different combination of atoms are considered, both coordinate-based PCA and LPH-based PCA are unable to identify the structure variation, as demonstrated in Figures \ref{fig:PCA_all_atom} and \ref{fig:PCA_all_atom_LPH} in the supplementary. In contrast, not only the general LWPH model works, we can also construct different LWPH by taking suitable combinations of backbone atoms and base atoms at local scale. For instance, according to CEHS scheme, we can take $C8$, $C4$, $N1$ and $C1'$ of purine base and $N3$, $C6$ and $C1'$ of pyrimidine base, as shown in Figure \ref{fig:base_pick}. These selected atoms based LWPH can also capture very well the confinement effect and ion-DNA configurational changes. The corresponding trajectories also show great consistence with both helical-parameter based and LWPH based results. Figures \ref{fig:PCA_LWPH_Select} and \ref{fig:MD_region_LWPH_CG} in the supplementary show the corresponding results.


Lastly, our LWPH based models are very flexible and easy to be combined with machine learning models. Traditional helical coordinate systems work well only for one and two base steps. They tend to fail if the local structure variation is cooperated between adjacent three or more base steps. Moreover, the helical coordinate systems are only suitable for DNA or RNA and cannot be used in proteins or other biomolecules. In comparison, our LWPH are more general and can be used for any local structures from DNAs, RNAs, proteins, biomocular complexes, or biomolecular assemblies. Another important property of LWPH based feature vectors are that they are unit free and can be used to compare different-sized local structures. In this way, these feature vectors are extremely suitable for machine learning models.

\section{Conclusion Remarks}\label{sec:conclusion}
In this paper, we discuss weighted persistent homology models and their applications in biomolecular structure, function, and dynamics analysis. We briefly review all the WPH approaches, including vertex-weighted, edge-weighted, and simplex-weighted models. Essentially, weight values, which reflects physical, chemical and biological properties, are assigned to vertices (atom centers), edges (bonds), or higher order simplexes (cluster of
atoms), depending on the biomolecular structure, function, and dynamics properties. Further, we propose the first localized persistent homology and localized weighted persistent homology and apply them in the DNA structure classification and clustering.

Our LPH and LWPH models are inspired by the great success of element specific persistent homology (ESPH), which does not treat biomolecules as an inseparable system, instead they are decomposed into a series of local domains, which may overlap with each other. The general persistent homology or
weighted persistent homology analysis is then applied on each of these local domains. In this way, functional properties, that embedded in our localized weighted homology, can be revealed. Our model has been used to systematically studying DNA structures

\section*{Acknowledgments}
This work was supported in part by Nanyang Technological University Startup Grant M4081842 and Singapore Ministry of Education Academic Research fund Tier 1 RG126/16 and RG31/18, Tier 2 MOE2018-T2-1-033.

\vspace{0.6cm}

\newpage

\section{Supplementary material}
\newpage

\begin{figure}
\begin{center}
\begin{tabular}{c}
\includegraphics[width=0.8\textwidth]{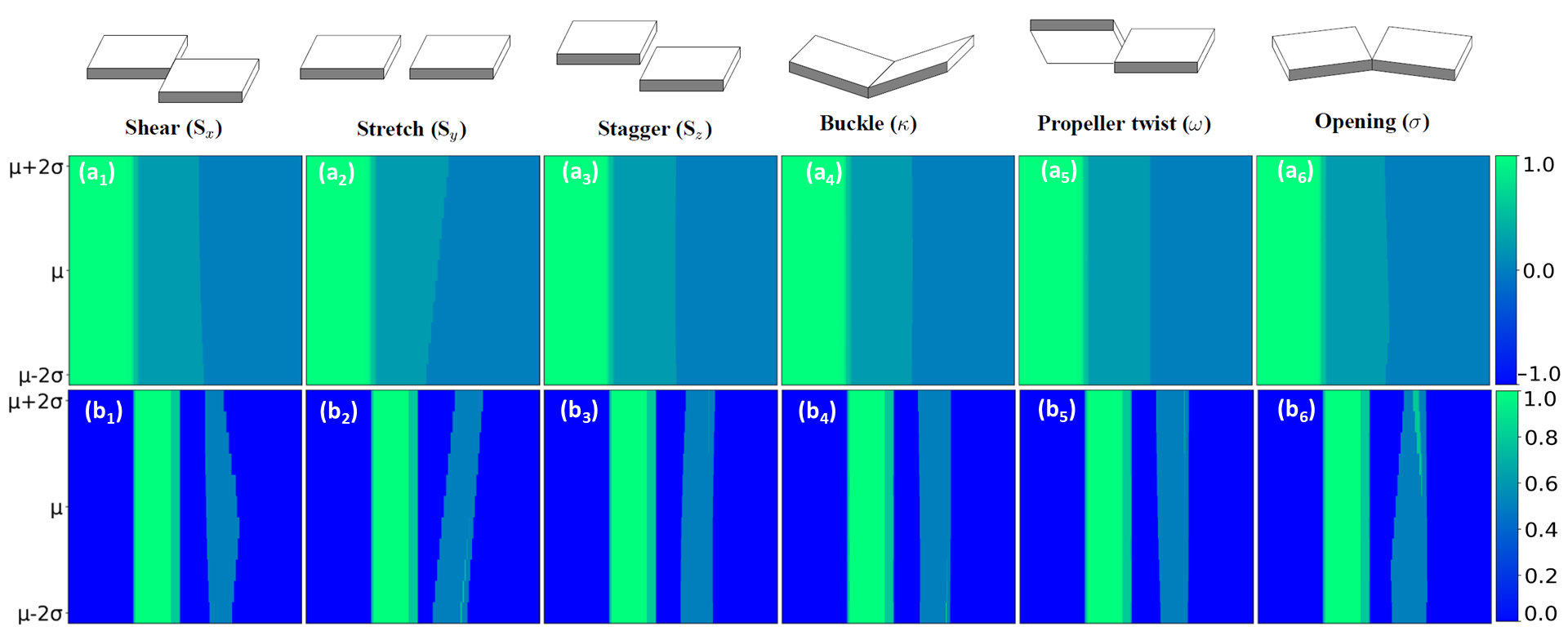}
\end{tabular}
\end{center}
\caption{The LPH based PBN image representation for each one-base-step helical parameter. In the $i$-th PBN image, we systematically change the $i$-th helical parameter value from  \(  \mu _{i}-2 \sigma _{i} \)  to  \(  \mu _{i}+2 \sigma _{i} \), with all other helical parameter remain as constant, to deliver a series of DNA structures. PBN can be calculated for each DNA structure and all of them stacked together to form a two-dimensional image. It can be seen that, both $\beta_0$ and $\beta_1$ PBN functions vary with the change of helical parameter value.
}
\label{fig:helical_onePair_orgM}
\end{figure}

\begin{figure}
\begin{center}
\begin{tabular}{c}
\includegraphics[width=0.8\textwidth]{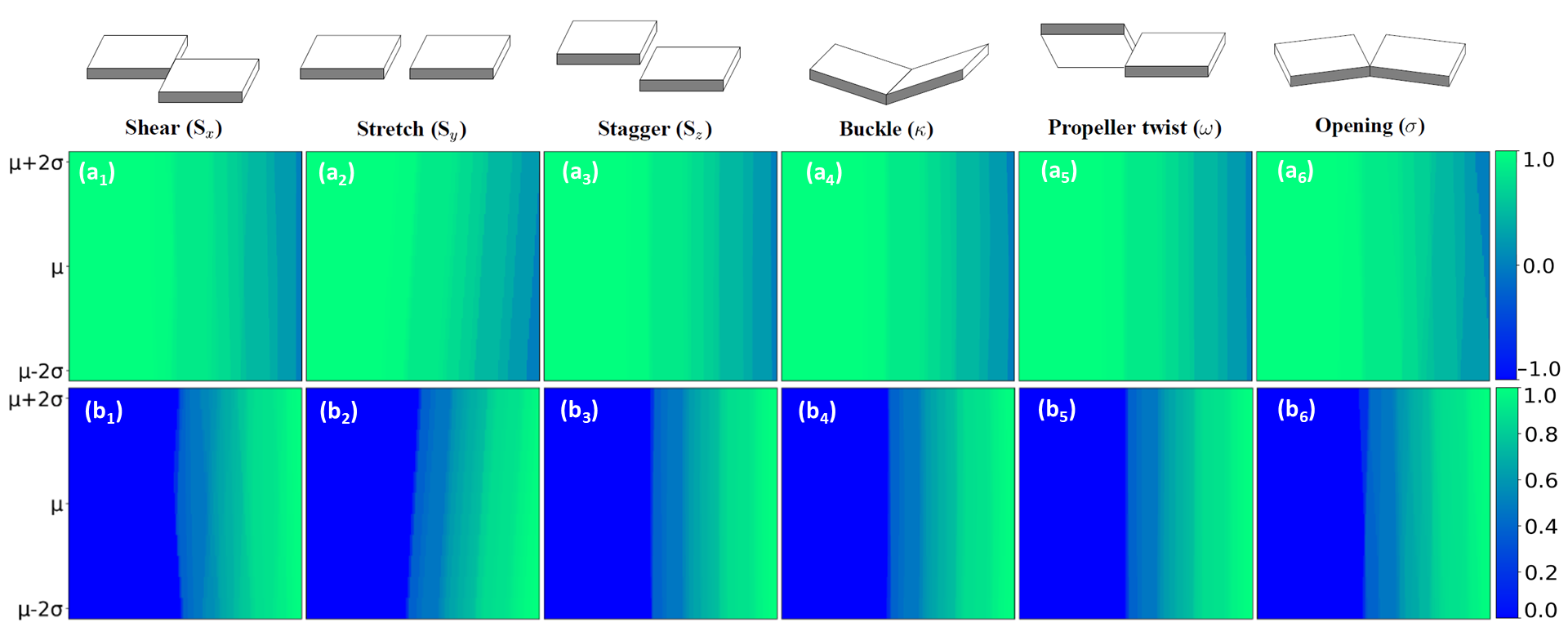}
\end{tabular}
\end{center}
\caption{The LWPH based PBN image representation for each one-base-step helical parameter. In the $i$-th PBN image, we systematically change the $i$-th helical parameter value from  \(  \mu _{i}-2 \sigma _{i} \)  to  \(  \mu _{i}+2 \sigma _{i} \), with all other helical parameter remain as constant, to deliver a series of DNA structures. PBN can be calculated for each DNA structure and all of them stacked together to form a two-dimensional image. It can be seen that, both $\beta_0$ and $\beta_1$ PBN functions vary with the change of helical parameter value.
}
\label{fig:helical_onePair_newM}
\end{figure}

\begin{figure}
\begin{center}
\begin{tabular}{c}
\includegraphics[width=0.95\textwidth]{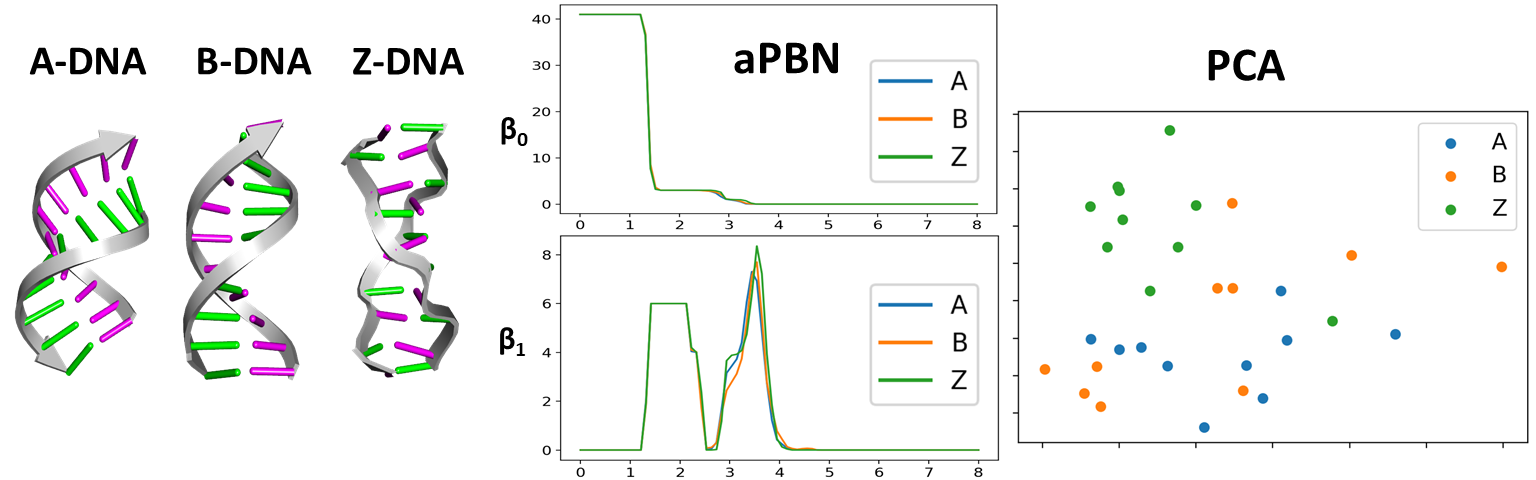}
\end{tabular}
\end{center}
\caption{LPH based classification of three DNA types, i.e., A-DNA, B-DNA, and Z-DNA. The average persistent Betti number (aPBN) from our LPH for three types of DNAs. We discretize the aPBN equally into a series of numbers and use these values as features for PCA. It can be seen that LPH based aPBN and PCA results cannot discriminate three DNA types. }
\label{fig:ABZ_LPH}
\end{figure}


\begin{table}
  \centering
	\caption{ The A-, B- and Z-types of proteins used in our paper. For each type, we arbitrary choose 10 PDB structures from the PDB databank.}
    \begin{tabular}{|c|c|c|} \hline
 A & B &Z\\ \hline
1DNZ &1BNA &1ICK  \\ \hline
1QPH &1D29 &1WOE  \\ \hline
2D47 &1D65 &1XAM  \\ \hline
2D94 &2L8Q &3P4J  \\ \hline
3V9D &2MCI &3WBO  \\ \hline
440D &3IXN &4HIG  \\ \hline
4IZQ &3U2N &4OCB  \\ \hline
5MVT &4C64 &5JZQ  \\ \hline
5WV7 &5T4W &6AQV  \\ \hline
5XK0 &6CQ3 &6BST  \\ \hline
    \end{tabular}
  \label{tab:PDBs}
\end{table}

\begin{figure}
\begin{center}
\begin{tabular}{c}
\includegraphics[width=0.9\textwidth]{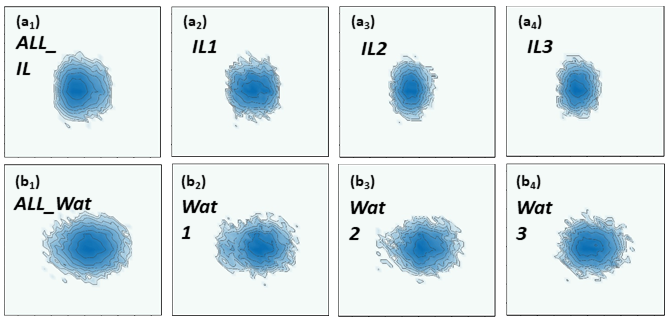}
\end{tabular}
\end{center}
\caption{The contour map generated from atom-coordinate PCA models from DNA configurations in IL and WAT environments. Similar to Figures (\ref{fig:PCA_LWPH}) and (\ref{fig:PCA_helical}), the x-axis and y-axis are the first and second principal components. Same notations for (${\bf a}_1$) to (${\bf a}_4$) and (${\bf b}_1$) to (${\bf b}_4$) are used. However, confinement effect and two center distribution of DNA configurations in IL environment are not observed.}
\label{fig:PCA_all_atom}
\end{figure}

\begin{figure}
\begin{center}
\begin{tabular}{c}
\includegraphics[width=0.9\textwidth]{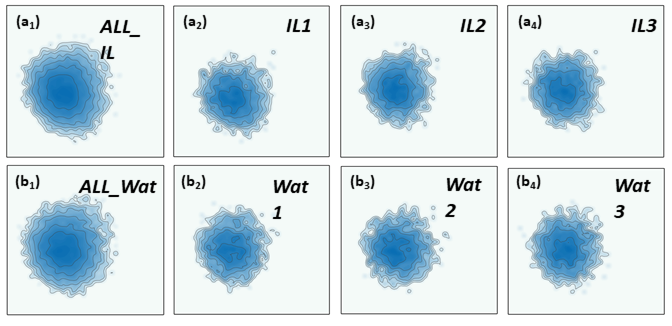}
\end{tabular}
\end{center}
\caption{The contour map generated from LPH based PCA models from DNA configurations in IL and WAT environments. Similar to Figures (\ref{fig:PCA_LWPH}) and (\ref{fig:PCA_helical}), the x-axis and y-axis are the first and second principal components. Same notations for (${\bf a}_1$) to (${\bf a}_4$) and (${\bf b}_1$) to (${\bf b}_4$) are used. However, confinement effect and two center distribution of DNA configurations in IL environment are not observed.}
\label{fig:PCA_all_atom_LPH}
\end{figure}

\begin{figure}
\begin{center}
\begin{tabular}{c}
\includegraphics[width=0.9\textwidth]{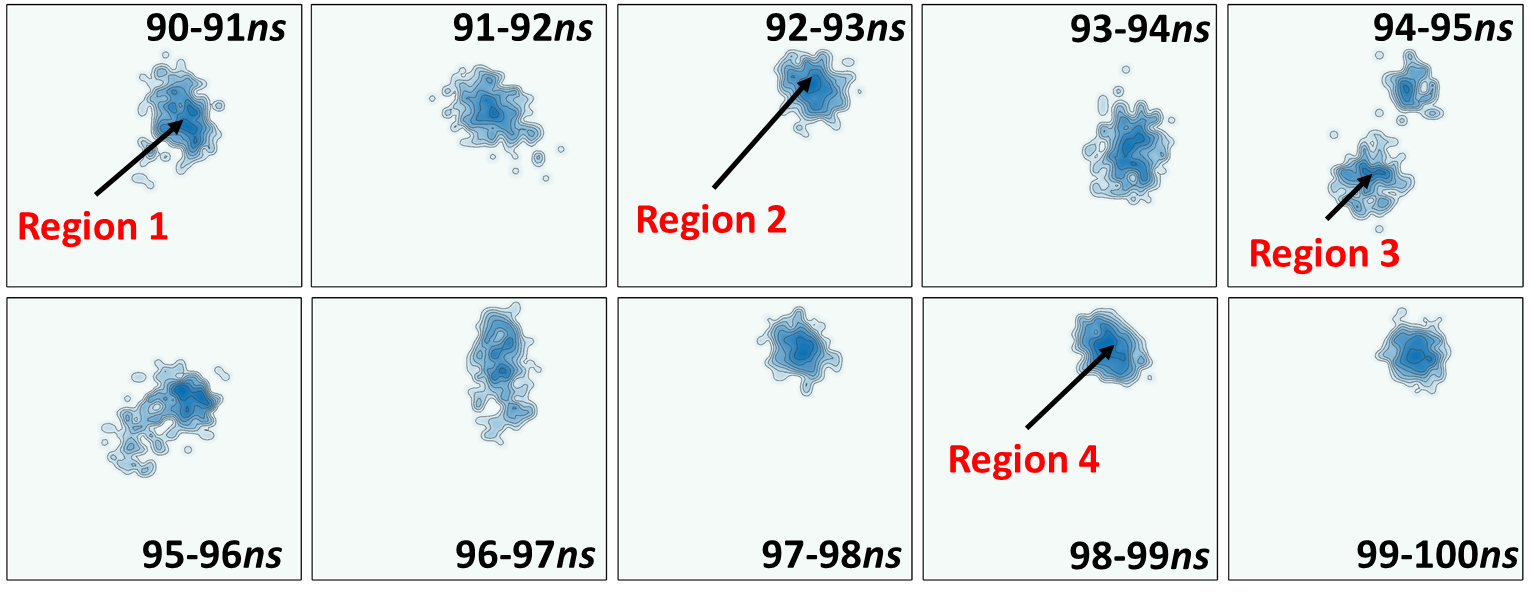}
\end{tabular}
\end{center}
\caption{The IL1 contour map generated from helical parameter based PCA models. Only the trajectories during the simulation time 90 to 100 nanoseconds are considered. Four different local regions can be identified, meaning that there are four different DNA configuration states. }
\label{fig:MD_region_HC}
\end{figure}

\begin{figure}
\begin{center}
\begin{tabular}{c}
\includegraphics[width=0.9\textwidth]{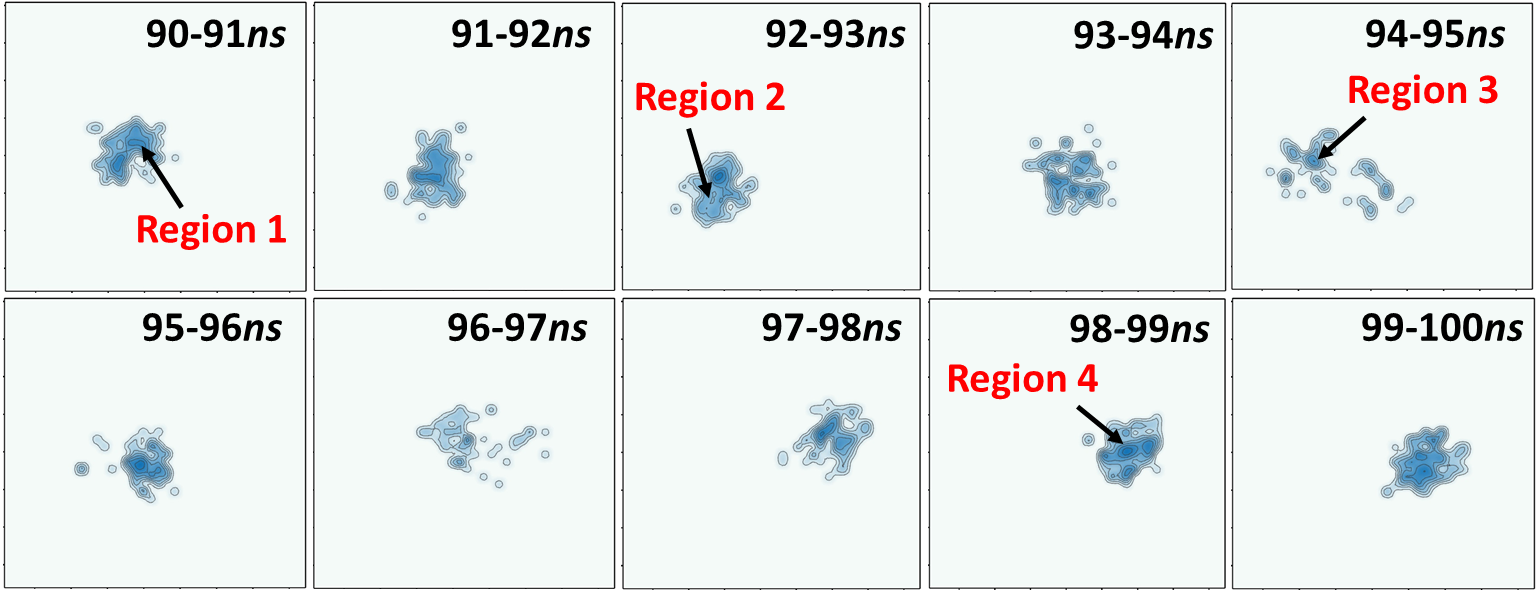}
\end{tabular}
\end{center}
\caption{The IL1 contour map generated from our LWPH based PCA models. Only the trajectories during the simulation time 90 to 100 nanoseconds are considered. Four different local regions can be identified, meaning that there are four different DNA configuration states. More importantly, these four states and the transition between the four states are highly consistent with the prediction from the helical parameter based models as in Figure \ref{fig:MD_region_HC}. }
\label{fig:MD_region_LWPH}
\end{figure}

\begin{figure}
\begin{center}
\begin{tabular}{c}
\includegraphics[width=0.9\textwidth]{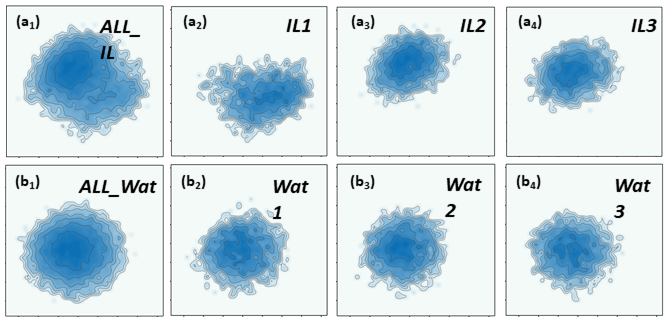}
\end{tabular}
\end{center}
\caption{The contour map generated from our LWPH based PCA models for selected atoms from DNA configurations in IL and WAT environments. Similar to Figures (\ref{fig:PCA_LWPH}) and (\ref{fig:PCA_helical}), the x-axis and y-axis are the first and second principal components. Same notations for (${\bf a}_1$) to (${\bf a}_4$) and (${\bf b}_1$) to (${\bf b}_4$) are used. Further, the same confinement effect and two center distribution of DNA configurations in IL environment as in Figures \ref{fig:PCA_LWPH} and (\ref{fig:PCA_helical}) are observed.}
\label{fig:PCA_LWPH_Select}
\end{figure}

\begin{figure}
\begin{center}
\begin{tabular}{c}
\includegraphics[width=0.9\textwidth]{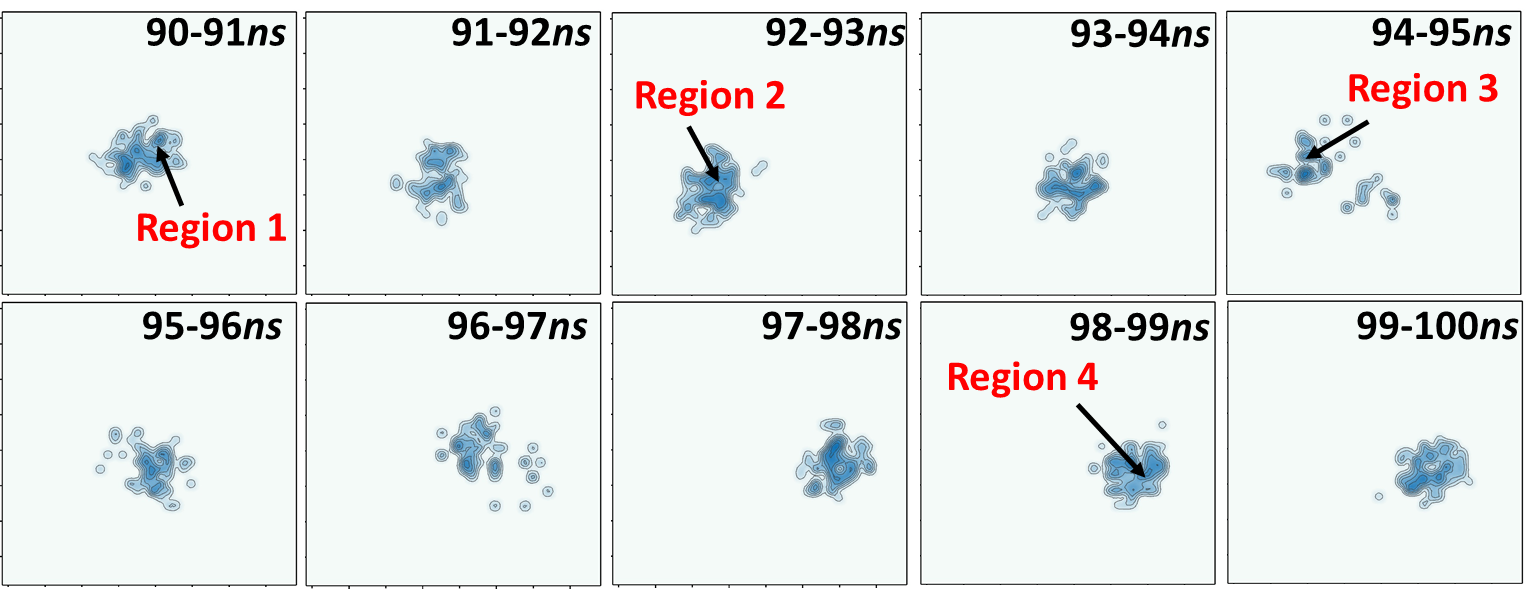}
\end{tabular}
\end{center}
\caption{The IL1 contour map generated from PCA of our selected atom based LWPH. Only the trajectories during the simulation time 90 to 100 nanoseconds are considered. Four different local regions can be identified, meaning that there are four different DNA configuration states. More importantly, these four states and the transition between the four states are highly consistent with the prediction from the helical parameter based models as in Figures \ref{fig:MD_region_HC} and \ref{fig:MD_region_LWPH} . }
\label{fig:MD_region_LWPH_CG}
\end{figure}

\end{document}